\documentclass[a4paper,11pt]{article}
\pdfoutput=1 

\usepackage{siunitx}
\usepackage{booktabs}
\usepackage{jinstpub}
\usepackage[table,xcdraw]{xcolor}
\usepackage{graphicx}
\usepackage{adjustbox}
\usepackage{multirow}
\usepackage{caption}
\usepackage{subcaption}
\title{\boldmath Intel Stratix 10 FPGA design for track reconstruction for the ATLAS experiment at the HL-LHC}

\author[a,1]{A. Camplani,\note{Corresponding authors.}}
\author[b,1]{S. Dittmeier,}

\author[c]{A. Annovi,}
\author[d]{K. Axiotis,}
\author[c]{R. Beccherle,}
\author[c]{N. Biesuz,}
\author[e]{R.~Brenner,}
\author[d]{S. Débieux,} 
\author[e]{M. Ellert,}
\author[c]{P. Francavilla,}
\author[c]{P. Giannetti,}
\author[f]{K. Kordas,}
\author[e]{M.~Mårtensson,} 
\author[c]{P. Mastrandrea,}
\author[f]{C. Noulas,}
\author[a]{J. Oechsle,}
\author[c]{M. Piendibene,}
\author[d]{R. Poggi,}
\author[b]{A.~Schöning,} 
\author[d]{A. Sfyrla,}
\author[c,2]{C. L. Sotiropoulou,\note{now with Computer Aided Software Technologies (CAST), Inc.}}
\author[e]{J. Steentoft,}
\author[f]{T. Tsiakiris,}
\author[a]{S. Xella}
\author[b]{and J.~Zinßer.} 

\affiliation[a]{Niels Bohr Institute (NBI), University of Copenhagen}
\affiliation[b]{Physikalisches Institut, Ruprecht-Karls-Universtität Heidelberg}
\affiliation[c]{INFN Pisa and Universita' di Pisa, Dipartimento di Fisica}
\affiliation[d]{Departement de Physique Nucleaire et Corpusculaire, Universite de Geneve}
\affiliation[e]{Uppsala University}
\affiliation[f]{Aristotle University of Thessaloniki}

\emailAdd{alessandra.camplani@cern.ch}
\emailAdd{sebastian.dittmeier@cern.ch}

\abstract{
The fast reconstruction of charged particle tracks with high efficiency and track quality is an essential part of the online data selection for the ATLAS experiment at the High Luminosity LHC. 
Dedicated custom designed hardware boards and software simulations have been developed to assess the feasibility of a Hardware Tracking Trigger (HTT) system.
The Pattern Recognition Mezzanine (PRM), as part of the HTT system, has been designed to recognize track candidates in silicon detectors with Associative Memory ASICs and to select and reconstruct tracks using linearized algorithms implemented in an Intel Stratix 10 MX FPGA.
The highly parallelized FPGA design makes extensive use of the integrated High-Bandwidth-Memory.

In this paper, the FPGA design for the PRM board is presented. Its functionalities have been verified in both simulations and hardware tests on an Intel Stratix 10 MX development kit.
}
\usepackage{lineno}

\keywords{Digital signal processing (DSP), Digital electronic circuits, Online farms and online filtering,  Trigger concepts and systems (hardware and software), Trigger algorithms, Pattern recognition, cluster finding, calibration and fitting methods}

\begin{document}
\maketitle

\flushbottom




\section{HL-LHC and the ATLAS experiment}
\label{sec:intro}
The High Luminosity Large Hadron Collider (HL-LHC)~\cite{Apollinari:2284929} is currently planned to begin operation in 2029. 
Protons will collide at $\sqrt{s}~=~\SI{14}{TeV}$, delivering an instantaneous luminosity of up to $\mathcal{L}~=~\SI{7.5e34}{cm^{-2}s^{-1}}$. 
This implies that the mean number of proton-proton collisions in the same bunch crossing (pile-up) is expected to reach up to $<\!\mu\!>\,=\,200$, an increase of more than a factor three compared to current conditions.
In order to achieve optimal physics performance under these harsh conditions, all LHC experiments will undergo a series of upgrades.

\begin{figure}[h]
    \centering
    \includegraphics[width=.7\textwidth]{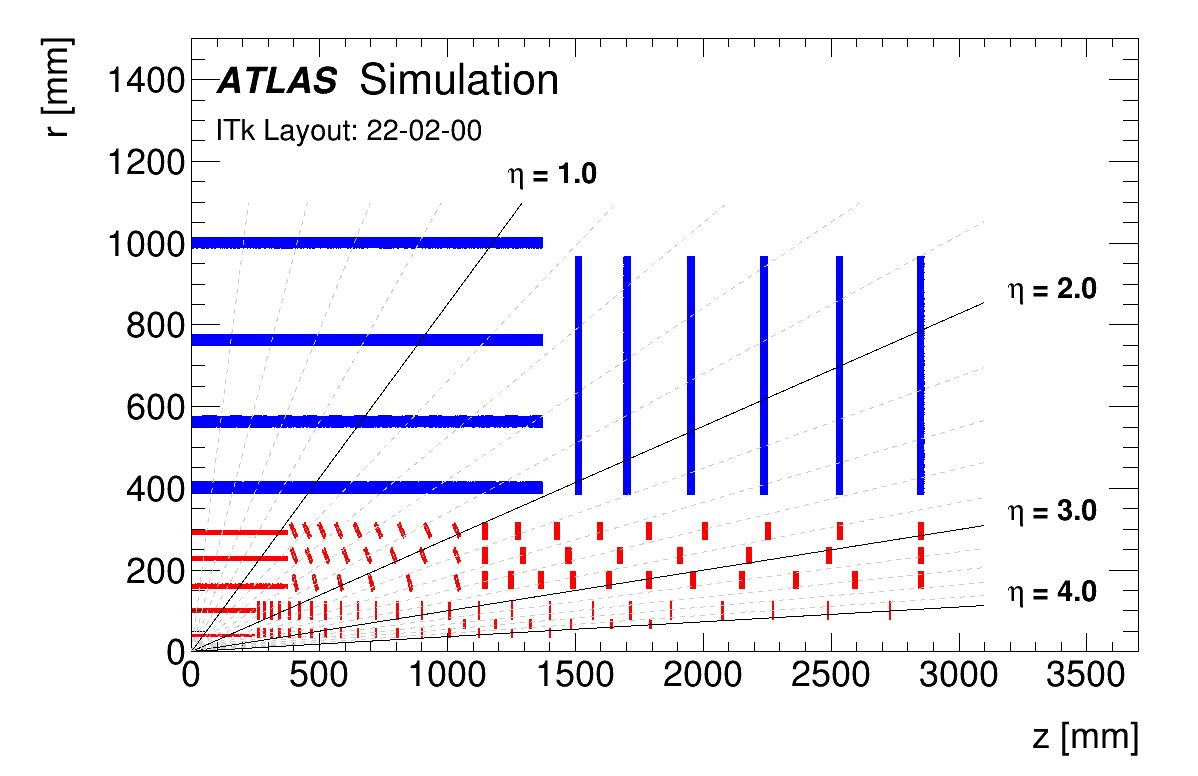}
    \caption{A schematic view of the layout of the ATLAS ITk detector used for this study, with a five-layer Pixel detector (red) surrounded by the Strip detector (blue). Only the positions of the active sensors are shown. The final ITk layout can be found in~\cite{atlas_itk}.}
    \label{fig:atlas_itk}
\end{figure}

For the HL-LHC era, the ATLAS experiment~\cite{ATLAS,ATLAS:LoI} foresees, among others, two major upgrades: the installation of a completely new tracking detector, the Inner Tracker (ITk), composed exclusively of silicon pixels~\cite{ITK:pix} and strips~\cite{ITK:str} as depicted in Figure~\ref{fig:atlas_itk}; and the redesign of the Trigger and Data Acquisition (TDAQ) system~\cite{TDAQP2}.

\subsection{The TDAQ Phase-II Upgrade}
\label{subsec:TDAQII}
The ATLAS detector currently produces more than \SI{60}{TB/s} of data during operation, but only a small fraction of these collision events contain interesting characteristics that might lead to new physics discoveries. 
As a consequence, a specialized multi-level system, the TDAQ system, selects the events that are potentially interesting for physics analyses in real-time. 
The TDAQ role is to reduce the data rates to manageable levels in order to be able to transport the data to the storage system for offline analysis.

At the HL-LHC, the ATLAS TDAQ system will face various challenges, such as a much more complicated event selection as a result of higher pile-up; readout of a tracking detector with more than 5 billion channels (compared to the $\sim$\num{100} million channels of the current detector); and a full event size of up to $\sim$\SI{5.2}{MB} (compared to the current $\sim$\SI{2}{MB}).

In addition, it is paramount for the trigger selection to remain inclusive enough to allow for electroweak scale physics as well as a broad range of physics searches.
Therefore, the TDAQ system will undergo a major upgrade, with several operating parameters updated with respect to the LHC era. Among these the most relevant are:
\vspace{-2.5mm}
\begin{itemize}
  \setlength\itemsep{-1mm}
  \item the hardware trigger latency will be increased from \SI{2.5}{\mu s} to \SI{10}{\mu s}
  \item the detector readout rate will be raised from \SI{100}{kHz} to \SI{1}{MHz}
  \item the rate of data sent to permanent storage will be increased from \SI{1}{kHz} to \SI{10}{kHz}
\end{itemize}
\vspace{-2.5mm}

Track reconstruction is a crucial element in the trigger selection, as it can provide extremely precise measurements in both position and momentum. This is of particular importance in order to distinguish hard-scattering processes from the abundant low-momentum background particles from concurrent pile-up events.

\begin{figure}[h]
 \centering
 \includegraphics[width=0.475\textwidth]
 {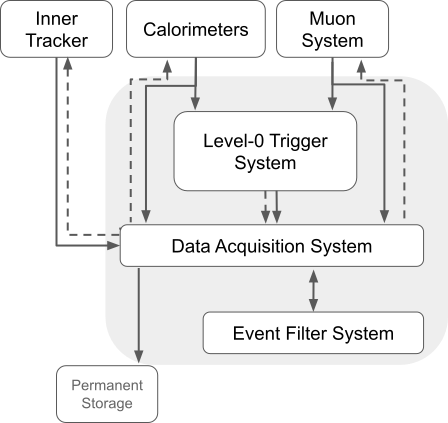}
 \caption{TDAQ system structure for Phase II. The dashed lines represent the Level-0 accept signal, while the solid lines represent the readout data.}
 \label{fig:tdqasys}
\end{figure}

The trigger system is composed of two levels: the hardware based Level-0 Trigger system, which relies only on data from the calorimeters and muon detectors and is operated with a fixed latency, and the software based Event Filter system, potentially deploying accelerator hardware, that provides high-level trigger functionality. Both trigger systems are connected to a complex Data Acquisition system. A schematic view can be found in Figure~\ref{fig:tdqasys}.

The Event Filter design in the Technical Design Report (TDR) \cite{TDAQP2} includes a CPU-based processing farm complemented by a hardware-based co-processor system, called HTT (Hardware Tracking for the Trigger).
The HTT system offers several benefits over a CPU-only farm, e.\,g.\, a lower processing latency and a more efficient use of resources.
The latter allows for a reduction of the CPU farm size required for online tracking by a factor of 10~\cite{TDAQP2}, which in turn leads to a lower power consumption of the overall system. 
This TDAQ architecture is referred to as the baseline design for the TDAQ upgrade, as in Figure \ref{fig:tdqasys}.

Given the availability of a hardware based tracking system, the TDR includes a scenario which evolves the trigger into a dual-level hardware trigger architecture.
This is done by introducing tracking at Level-1 using a dedicated low-latency version of HTT.
This can reduce the L0 rate to less than \SI{1}{MHz} for further processing in the Event Filter.
Implementing hardware tracking at Level-1 enables better control of the trigger rates by adding more margins and more flexibility, not only in the case of major changes of the experimental conditions or the trigger menu, but also as a handle to maintain stability as the processing at high pile-up does not scale linearly with luminosity.
Furthermore, it creates opportunities to lower thresholds and improve acceptance on many important physics signatures.


As the HL-LHC upgrade projects progressed, the final choice of the ATLAS experiment was to design a heterogeneous system, based on commodity servers and possibly accelerator boards, excluding the possibility of a system based around custom associative memory ASICs \cite{TDAQ_amend}. Nevertheless, the decision process required performance studies, both in simulation and on hardware demonstrators, to assess the feasibility of such a system.
In this paper we present the development of the firmware design for one mezzanine board dedicated to the track reconstruction and fitting, called the Pattern Recognition Mezzanine.

\subsection{The Hardware Tracking for the Trigger system}
\label{sec:htt}
The HTT system is designed to reconstruct quickly and efficiently the tracks of charged particles within the ITk.
Depending on the Level-0 trigger decision, the ITk detector data can be processed in two different ways in the baseline HTT design: regional or full-scan track reconstruction, according to trigger configuration.

\begin{table}[h]
\centering
\begin{tabular}{lccc}
\toprule
\multicolumn{1}{c}{\textbf{}} & \multicolumn{1}{c}{\textbf{$p_{\rm{T}}$}} & \textbf{\begin{tabular}[c]{@{}c@{}}Detector\\ coverage\end{tabular}} & \textbf{Level-0 rate} \\
\midrule
\textbf{Regional tracking} & \textgreater \SI{2}{GeV} & \textless{} \SI{10}{\%} & \SI{1}{MHz} \\
\textbf{Full-scan tracking} & \textgreater \SI{1}{GeV} & full coverage & \SI{100}{kHz} \\
\bottomrule
\end{tabular}
\caption{Regional and full-scan tracking requirements.}
\label{tab:reg_full}
\end{table}

For the regional tracking, as shown in Table \ref{tab:reg_full}, tracks with transverse momentum of $p_{\rm{T}} > \SI{2}{\GeV}$ are reconstructed within the so-called Regions Of Interest (ROI), which cover less than \SI{10}{\%} of the detector volume.
This allows for a quick initial background rejection for single high-$p_{\rm{T}}$ lepton triggers, as well as for multi-object triggers.
Regional tracking can be done at the full Level-0 accept rate of \SI{1}{MHz}.

For the full-scan reconstruction, all tracks with $p_{\rm{T}} > \SI{1}{\GeV}$ are reconstructed using the full detector data.
This is especially important for background rejection of hadronic signatures, to enhance primary vertex identification, and to correct and mitigate pile-up effects.
As these requests demand more bandwidth and processing power, they can be served at rates of up to \SI{100}{kHz}.

For the scenario where the Level-1 Trigger is present, the low-latency version of the HTT system has to provide tracks to the next processing step within a few microseconds. In this case only the regional tracking is used.
For the HTT system placed in the Event Filter, both regional and global tracking are available, as in the baseline design.

The requirements posed by the low-latency version are the main driver for the architectural choice of the HTT system.
The high throughput and the tight latency budget can only be met using a dedicated hardware system.
FPGAs that can sustain the required throughput have been selected to implement the clustering and track fitting algorithms.
A custom-designed ASIC has been developed, the Associative Memory (AM) ASIC (described in detail in~\cite{AM08}), following the approach previously used by the CDF experiment~\cite{CDF:2003mka} and intended for the ATLAS FastTracKer (FTK) project~\cite{ATLAS:2021tfo}. 
The AM ASICs are able to deal with the combinatorial complexity in an extremely parallel way, providing a fast recognition of track candidates with high efficiency.
In parallel to the developments of the AM ASICs, hardware implementations of alternative pattern recognition methods using Hough transform have also been studied~\cite{Martensson}.
Related studies of non-AM based track triggers can be found in~\cite{Aggleton,CMS_fitter}.

\begin{figure}[h]
 \centering
 \includegraphics[width=0.4\textwidth]{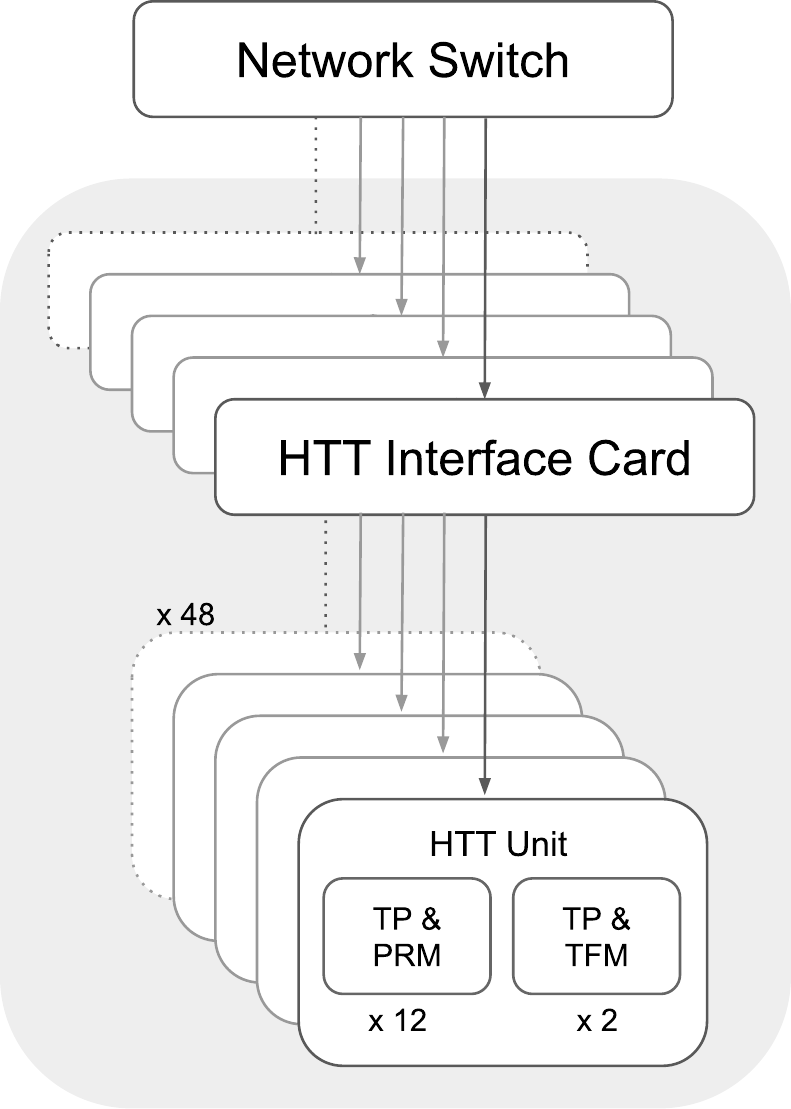}
 \caption{HTT system architecture.}
 \label{fig:httsys}
\end{figure}

The HTT design relies on a single ATCA \cite{ATCA} main board, the Tracking Processor (TP), that can be equipped with two different types of mezzanines: the Pattern Recognition Mezzanine (PRM) or the Track Fitting Mezzanine (TFM). 
The TP mounts a Xilinx Virtex UltraScale+ VU19P FPGA.
The PRM and the TFM both host an Intel Stratix 10 MX FPGA. 
The AM ASICs are only mounted on the PRM.
In the baseline design, see Figure~\ref{fig:httsys}, 12 PRM boards and 2 TFM boards share one ATCA crate and form one HTT unit. 
The detector volume is split into 48 $\eta \times \phi$ regions ($2.66 \times \pi/8$), each being individually processed by one HTT unit.
The segmentation has been chosen to minimize data sharing between HTT units required to account for regional overlaps. 

The HTT system receives the detector data via optical links from the network through dedicated servers using the FELIX~\cite{felix_inTDR, felix_proc} cards. 
In the Level-1 case, dedicated low-latency streams are necessary to transfer the data directly from the ITk to the regional tracking units.

The HTT system processes the ITk detector data in steps.
The incoming ITk hits are clustered on the TP and clusters are sent to the mezzanines. 
The PRM reconstructs tracks from 8 ITk layers, while the TFM refines the fit adding the hits from the remaining five layers.

\section{The Pattern Recognition Mezzanine}
\label{sec:prm}

\begin{figure}[h]
 \centering
 \includegraphics[width=0.6\textwidth]{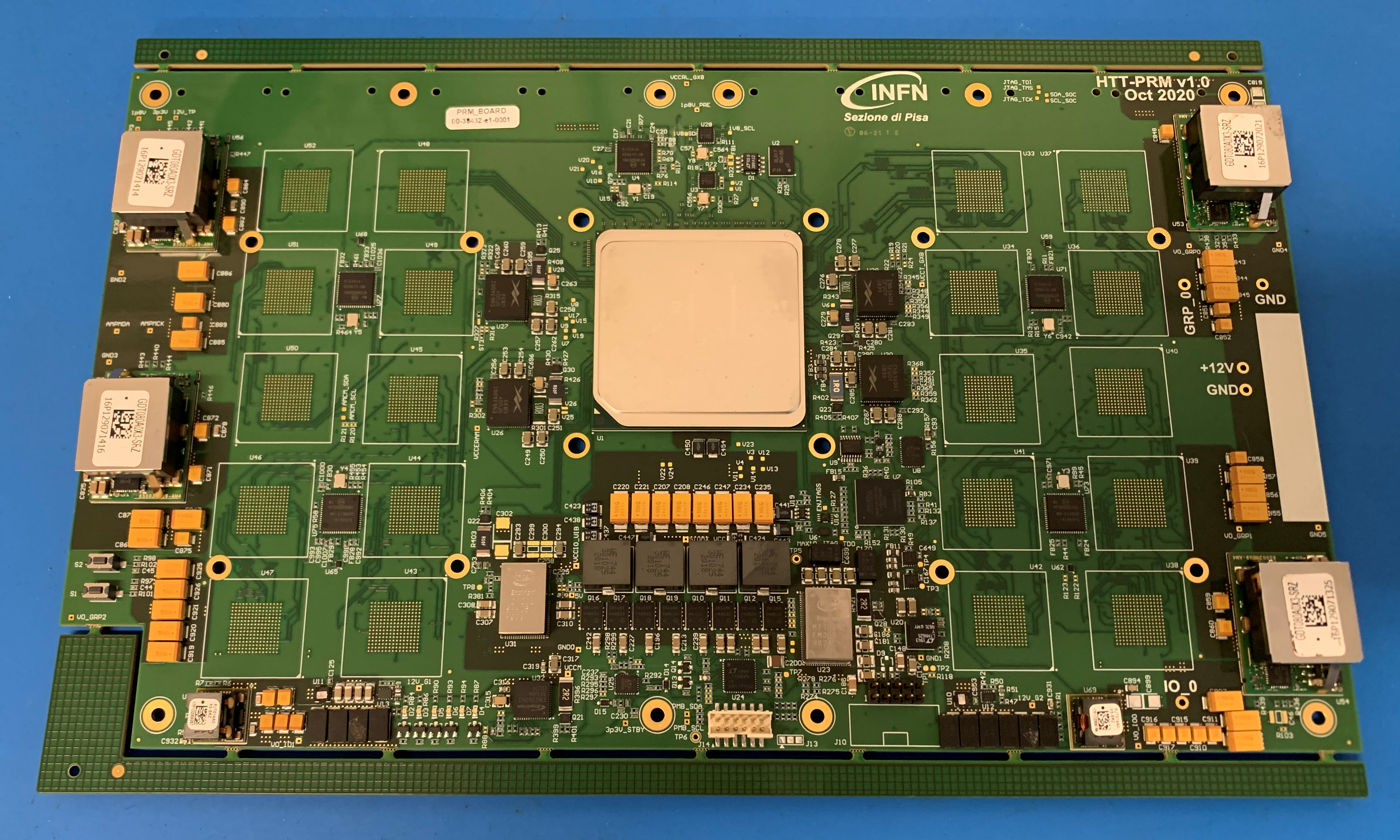}
 \caption{Photograph of a Pattern Recognition Mezzanine board equipped with an Intel Stratix 10 MX FPGA. No AM ASICs are soldered to the board.}
 \label{fig:prmboard}
\end{figure}

A prototype of the Pattern Recognition Mezzanine (PRM) is shown in Figure~\ref{fig:prmboard}.
The PRM PCB is meant to be mounted onto its ATCA motherboard (TP), and it features a powerful FPGA, the Intel Stratix 10 MX\footnote{part number: 1SM21BHN3F53E3VG}, together with 20 Associative Memory ASICs.
To implement the power-up sequence of the Stratix 10 as well as for control and monitoring of the board, an Intel MAX 10 FPGA is also installed.
The PRM connects to the TP via eight Samtec Z-Ray connectors of type \mbox{ZA8H-24-0.33-7}, that provide power and high speed serial data links, as well as SGMII~\cite{SGMII}, JTAG~\cite{JTAG} and other control interfaces.

\subsection{Functional description}
The PRM performs the first track reconstruction and selection step for the HTT system.
It processes data from 8 detector layers, of which a maximum of four can be silicon pixels and the rest are silicon strip layers, depending on which ITk detector region is being processed \cite{ITK:pix,ITK:str}.
As in the HTT simulation studies, the configuration used for this study foresees 1 pixel and 7 strip layers.
The Intel Stratix 10 MX FPGA forms the core of the PRM, handling the communication with the TP and the ASICs as well as implementing the algorithms used for track reconstruction and selection.
The PRM receives clustered ITk detector data via high speed serial links from the TP using the FPGA's transceiver blocks. 
Up to 32 channels for the input and 16 for the output data have been foreseen.

\begin{figure}[h]
 \centering
 \includegraphics[width=0.55\textwidth]{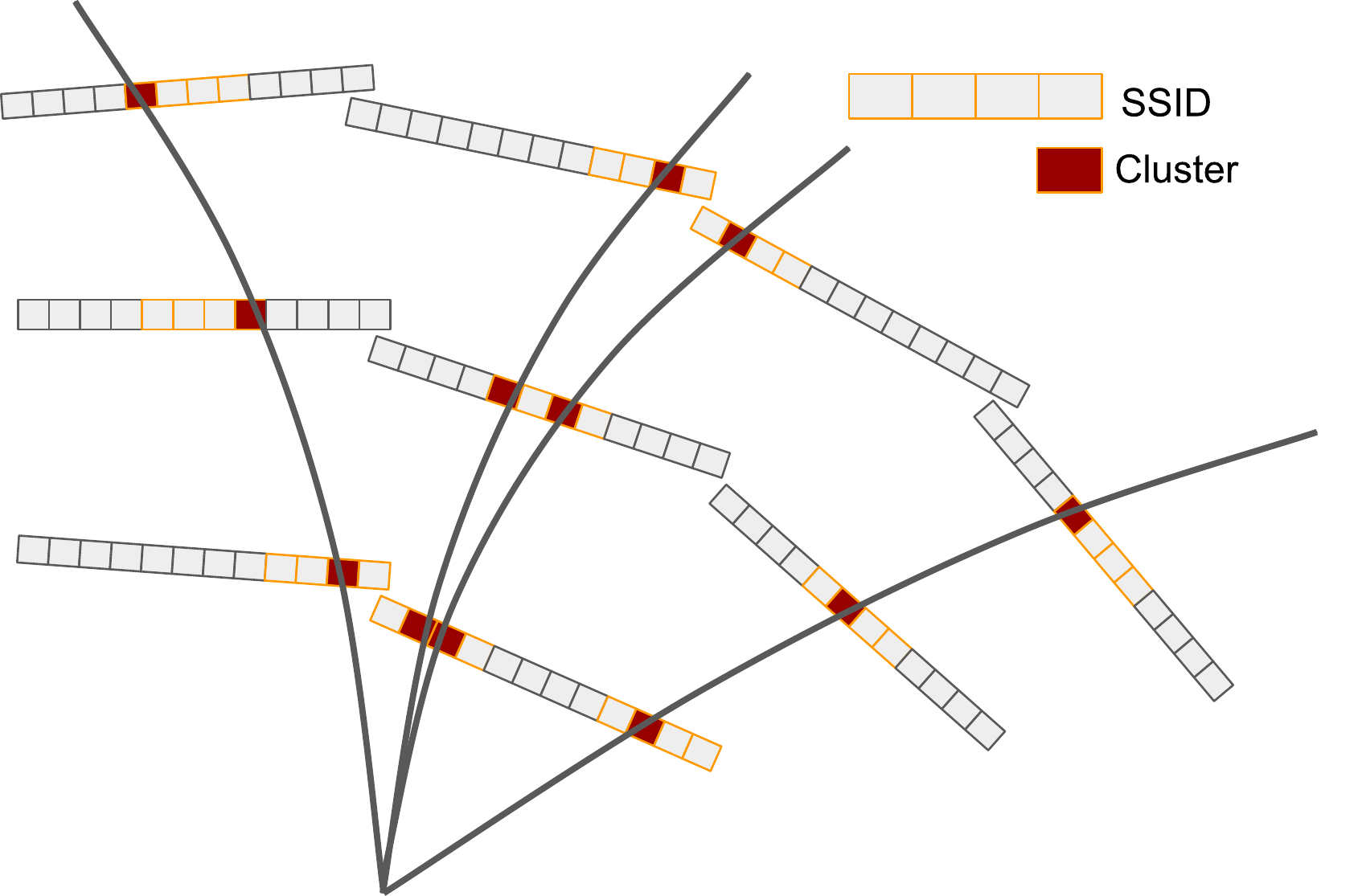}
 \caption{Simplified representation of super-strips identifiers granularity vs. single clusters granularity: one super-strip can contain multiple clusters.}
 \label{fig:cl_ssid}
\end{figure}

Twenty AM ASICs on the PRM are arranged in four groups of five daisy-chained ASICs, that can process different sets of clusters simultaneously.
All 20 ASICs combined can store a total of 7.864 million patterns.
A pattern is a sequence of eight so-called Super-Strip IDentifiers (SSIDs), which collect adjacent groups of silicon strip or pixel channels into a coarse position identification. Figure~\ref{fig:cl_ssid} shows a simplified representation\footnote{HTT simulation studies have used the following SSID sizes: 33x402 for the pixel and 40 for the strips in the barrel, and 16x200 for the pixel and 10 for the strips in the end-caps.}.

All the patterns are generated from Monte Carlo simulated muons\footnote{Efficient tracking of long living particles, i.\,e.\,largely displaced tracks, requires additional patterns and constants to be generated. 
Similarly, tracking efficiency of lower energy electrons can be retained with additional patterns, if no further processing, like a Gaussian Sum Filter, is available.}, which originate from the interaction region in ATLAS with a flat $p_{\rm{T}}$ spectrum for $p_{\rm{T}}$ larger than \SIrange{1}{2}{GeV} \cite{Sim-Seba}.
The super-strip width, in both dimensions for the pixels, is a parameter tuned in the simulation to optimize the number of patterns within the AM ASICs and to reduce the bandwidth between the FPGA and ASICs. 
At the same time, it impacts on the average number of fit combinations, as an increase in the super-strip width would also increase the combinatorics.

\begin{figure}[h]
 \centering
 \includegraphics[width=0.9\textwidth]{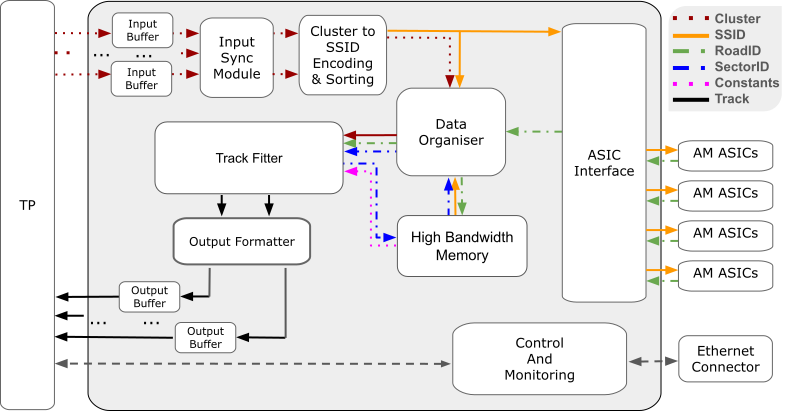}
 \caption{PRM complete block diagram. Solid lines represent the data transmission: in red the clusters, in orange the SSID, in green the roadID, in blue the sectorID, in violet the constants and in black the tracks. Dashed lines represent the firmware control and monitoring.}
 \label{fig:s10_complete}
\end{figure}

The complete block diagram of the Intel Stratix 10 MX FPGA design functionalities is shown in Figure \ref{fig:s10_complete} and explained in the following paragraphs.
The data processing is event based, where one event corresponds to the collision of two proton bunches.
In the first processing step, the input cluster data (red) coming from the TP is synchronized and converted into SSIDs (orange), which are sent to the AM ASICs for pattern matching.
The interface between the FPGA and a group of ASICs consists of source-synchronous parallel LVDS buses operated at \SI{1}{Gbps} per line, which yields an ASIC input data rate of \SI{32}{Gbps} and an ASIC output data rate of \SI{8}{Gbps}.

Within the FPGA, all clusters are stored temporarily in an on-the-fly data base called Data Organiser, grouped under their corresponding SSIDs.
When an event processing is being completed in the AM ASICs, the FPGA receives the addresses of all matched patterns, which are referred to as road IDentifiers (roadIDs, green).
Using these roadIDs, the SSIDs of all matched patterns are collected in the Data Organizer from the High-Bandwidth-Memory (HBM), where a copy of the pattern bank is stored.
The SSIDs and their associated clusters are then processed in the Track Fitter block.

The Track Fitter block implements a Principal Component Analysis (PCA) algorithm to determine the track parameters for a given set of clusters.
A specific PCA is only applicable within a small detector region, referred to as a sector.
The detector is therefore divided into many sectors which are identified by their sectorIDs (blue), stored in the HBM.
Within a sector the track equation can then be linearized, according to equation~\ref{eq:param}.

\begin{align}
\label{eq:param}
p_i &= \sum^{N_{\text{coo}}}_{j=1} \left( C_{ij} x_{j}  + q_{i} \right)
\end{align}

Here, $x_j$ are the coordinates of the clusters and $p_i$ are the 5 expected track parameters, namely pseudorapidity $\eta$, azimuthal angle $\phi$, transverse momentum $p_{\rm{T}}$, and the impact parameters $d_0$ and $z_0$. 
$C_{ij}$ and $q_i$ form the set of constants extracted from the software simulation training which are stored in the HBM (violet). 
The minimization of the distance between the expected and real track parameters in the PCA follows a $\chi^2$-distribution, allowing the use of the $\chi^2$-variable to evaluate the track quality. 

\begin{align}
\label{eq:chi}
\chi^{2} &= \sum^{N_{\text{dof}}}_{i=1} \left[ \left( \sum^{N_{\text{coo}}}_{j=1} S_{ij} x_{j} \right) + h_{i} \right]^{2}
\end{align}

Equation~\ref{eq:chi} contains the constants $h_{i}$ and $S_{ij}$, where $S_{ij}$ is the inverted covariance matrix between the coordinates, also stored in the HBM (violet).
The number of coordinates is $N_{\text{coo}}~=~2N_{\text{pixels}}~+~N_{\text{strips}}$, with pixels contributing two coordinates and strips only one. 
$N_{\text{dof}}$ is the number of degrees of freedom, $N_{\text{dof}}~=~N_{\text{coo}}~-~N_{\text{par}}$, with the number of track parameters $N_{\text{par}}~=~5$.
It should be noted that the algorithm also allows for tracks with up to two missing clusters, e.\,g.\, due to detector inefficiencies.
In these cases, the track parameters and $\chi^2$ can be recovered by a minimization with respect to the coordinates of the missing clusters, or by generating additional sets of constants that could be directly applied into equations~\ref{eq:param} and \ref{eq:chi}.



The computations in the Track Fitter block rely solely on multiply-accumulate (MAC) operations.
Hence, they can be efficiently implemented with the FPGA's integrated DSPs.
A cut on the $\chi^2$ value is implemented to reject most of the wrong combinations of clusters before computing the track parameters, such that only about \SI{8}{\%} of the tracks are selected, as shown in Table \ref{tab:PRM_parameters}.
This concludes the data processing on the Intel Stratix 10 FPGA.
For each track (black), its track parameters, its $\chi^2$ value, and the associated cluster information (represented by
the black lines) are sent from the PRM to the TP via high speed serial links.

\subsection{Requirements}
Table~\ref{tab:PRM_parameters} summarizes the requirements for the Pattern Recognition Mezzanine in the TDAQ baseline architecture with HTT as part of the Event Filter System.
These requirements are derived from simulations of the expected data flow through the entire HTT system for pile up <$\mu$> = 200 \cite{Sim-Seba}.
The main parameters that affect the processing speed are the average cluster rate, road processing rate, fit constants access rate, and track fitting rate.
In the low-latency scenario, an additional constraint on the processing latency of $\mathcal{O}(\SI{1}{\micro\second})$ has to be fulfilled.
The HBM storage requirements for a full PRM are about \SI{73}{MB} to hold the patterns, and about \SI{375}{MB} for the constants, estimated from simulation studies.  
Multiple copies of the patterns and constants can be stored in the HBM to fulfil the rate requirements. 

\begin{table}[h]
\centering
\caption{Main parameters for the PRM and their requirements derived from the HTT system in the TDAQ Phase-II baseline design. 
The values are related to an expected maximum event request rate of \SI{240}{kHz} and refer to one of four firmware units working in parallel within the PRM that are connected to separate groups of 5 AM ASICs.}
\label{tab:PRM_parameters}
\footnotesize{
\begin{tabular}{lr}
\toprule
\textbf{} & \textbf{Requirements}\\
\textbf{Parameter} & \textbf{per group of ASICs}\\
\midrule
Peak cluster rate per layer 					& \SI{95}{MHz}   \\
Roads processing rate 							& \SI{77}{MHz}    \\
Constants readout rate ($\chi^2$ calculation) 	& \SI{41}{MHz}    \\
Fit rate ($\chi^2$ calculation) 				& \SI{714}{MHz}   \\
Constants readout rate (parameter calculation) 	& \SI{41}{MHz}    \\
Fit rate (parameter calculation)                & \SI{57}{MHz}    \\
Track output rate 								& \SI{57}{MHz}    \\
\bottomrule
\end{tabular}}
\end{table}


\section{Block description of firmware for standalone testing}
\label{sec:blocks}

\begin{figure}[h]
 \centering
 \includegraphics[width=0.5\textwidth]{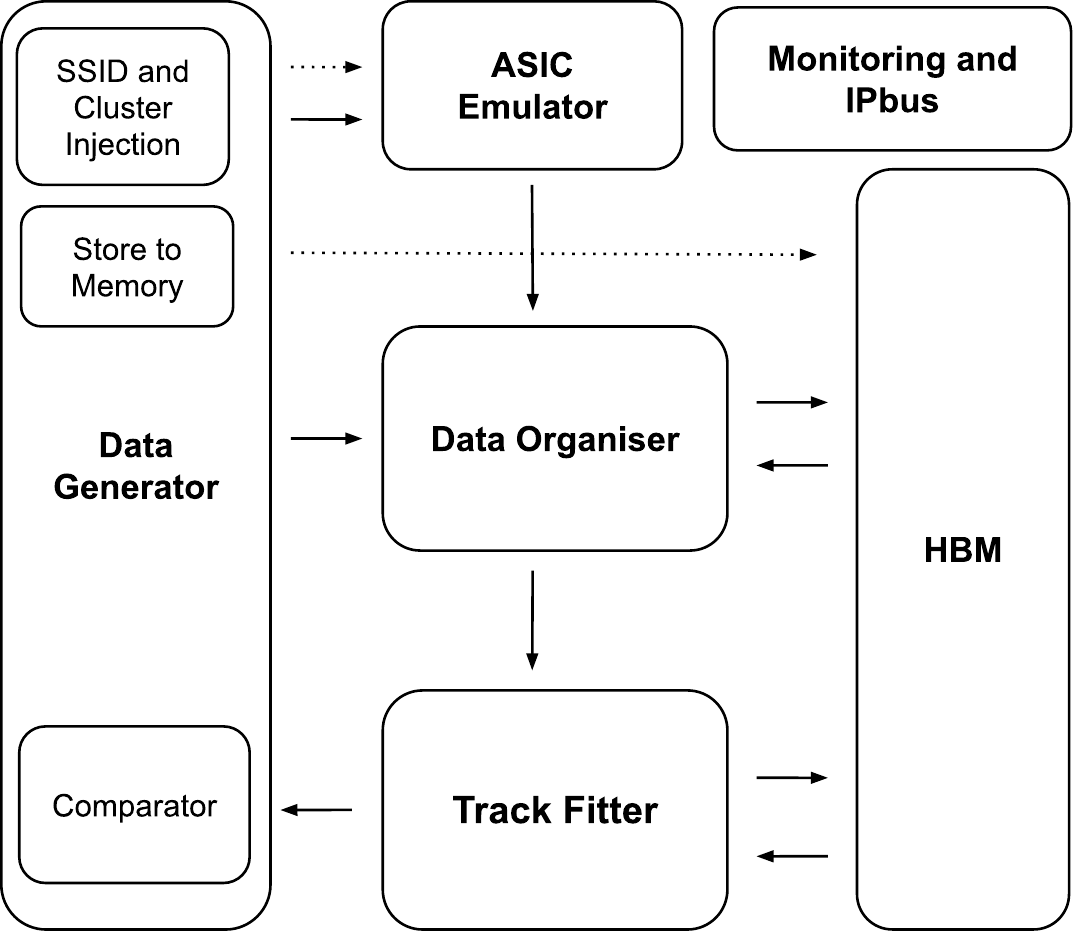}
 \caption{PRM standalone block diagram. The dashed arrows are for paths representing the storage of patterns or constants into memory. The solid line arrows are for data paths.
}
 \label{fig:s10_standalone}
\end{figure}

The PRM firmware has been developed and tested in a "standalone" set-up, where the transceivers and the ASIC functionalities have been emulated through a Data Generator and an ASIC Emulator block.
The corresponding firmware diagram is shown in Figure \ref{fig:s10_standalone}.
In the following, the individual blocks are described in more detail in the order of the data flow. 

\subsection{Data Generator}

\begin{figure}[h]
 \centering
 \includegraphics[width=0.8\textwidth]{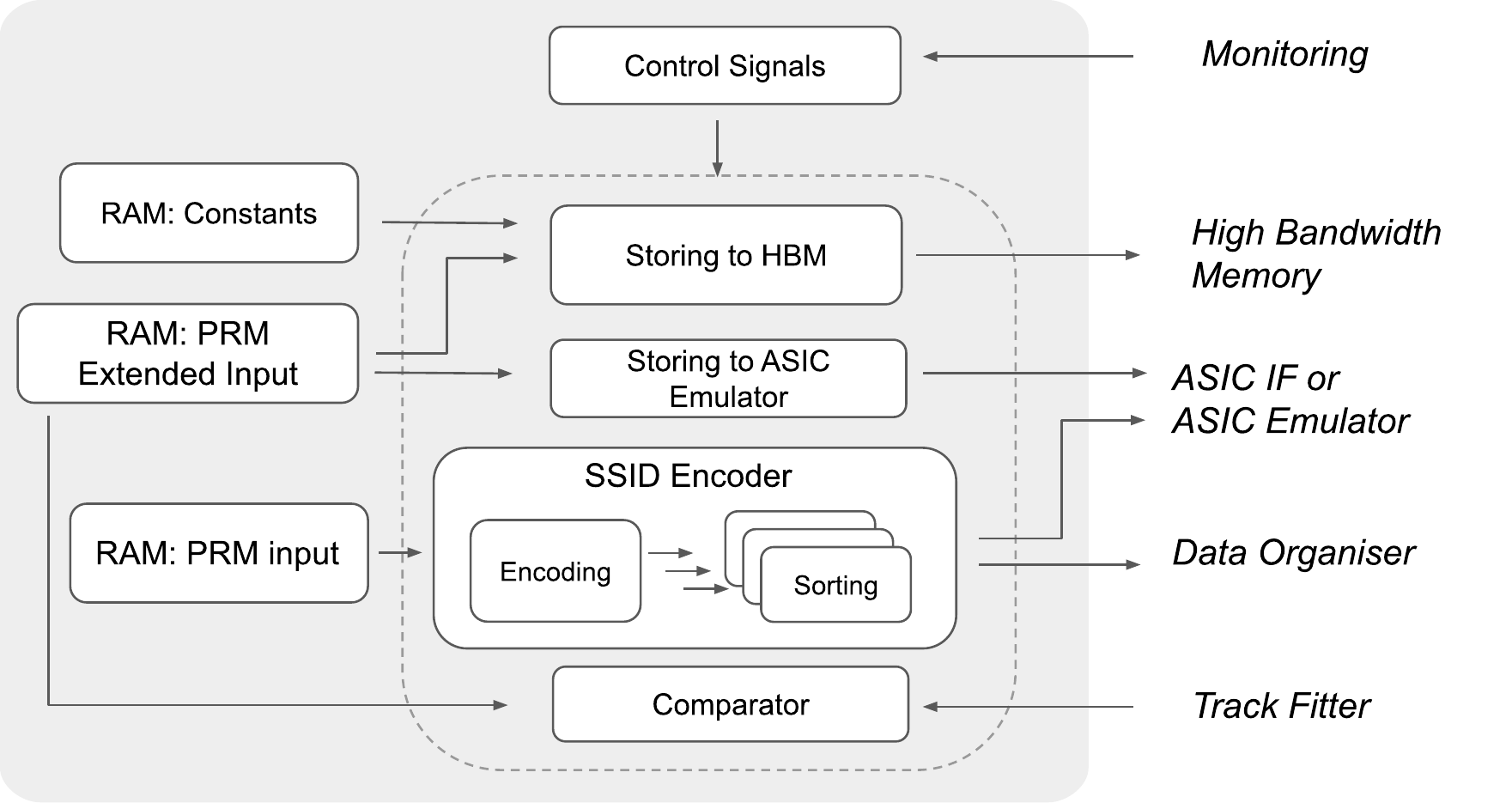}
 \caption{Data Generator block diagram with activated SSID Encoder.
 }
 \label{fig:data_generator_diagram}
\end{figure}

The Data Generator block role is to provide different test modes, initialize the PRM internal memories with the necessary constants, inject the data, and check the results. The block diagram is shown in Figure~\ref{fig:data_generator_diagram}.

For the first functionality, the Control Signal block handles the different testing modes.
Two injection modes have been implemented: with and also without the SSID Encoder block, responsible for the assignment of an SSID to each cluster. The selection between the two modes is done at compilation time.
In both modes various parameters of the data injection can be controlled, such as the injection frequency and the number of events.
In addition, it is possible to add randomly generated fake clusters and SSIDs into the data stream, to emulate the real detector behavior in the high-pile-up environment.


The PRM internal memories are initialized when a memory initialization command is issued. 
The fit constants are uploaded from the Constants RAM into the High-Bandwidth Memory, presented in Section \ref{sec:HBM}, through the Storing to HBM block.
The track patterns are also sent from the PRM Extended Input RAM to the ASIC Emulator block, described in Section \ref{sec:Emu}, and to the HBM via the Storing to ASIC Emulator and HBM blocks.

Following the initialization of the memories, the injection is started. The PRM Input RAM contains the clusters in the same format as expected to be computed in the TP motherboard.

For the results check, the PRM Extended Input RAM contains the simulated true track $\chi^2$ and the helix parameters for an easy comparison that happens in the Comparator block. 

\subsection{ASIC Emulator}
\label{sec:Emu}
\begin{figure}[ht]
 \centering
 \includegraphics[width=0.8\textwidth]{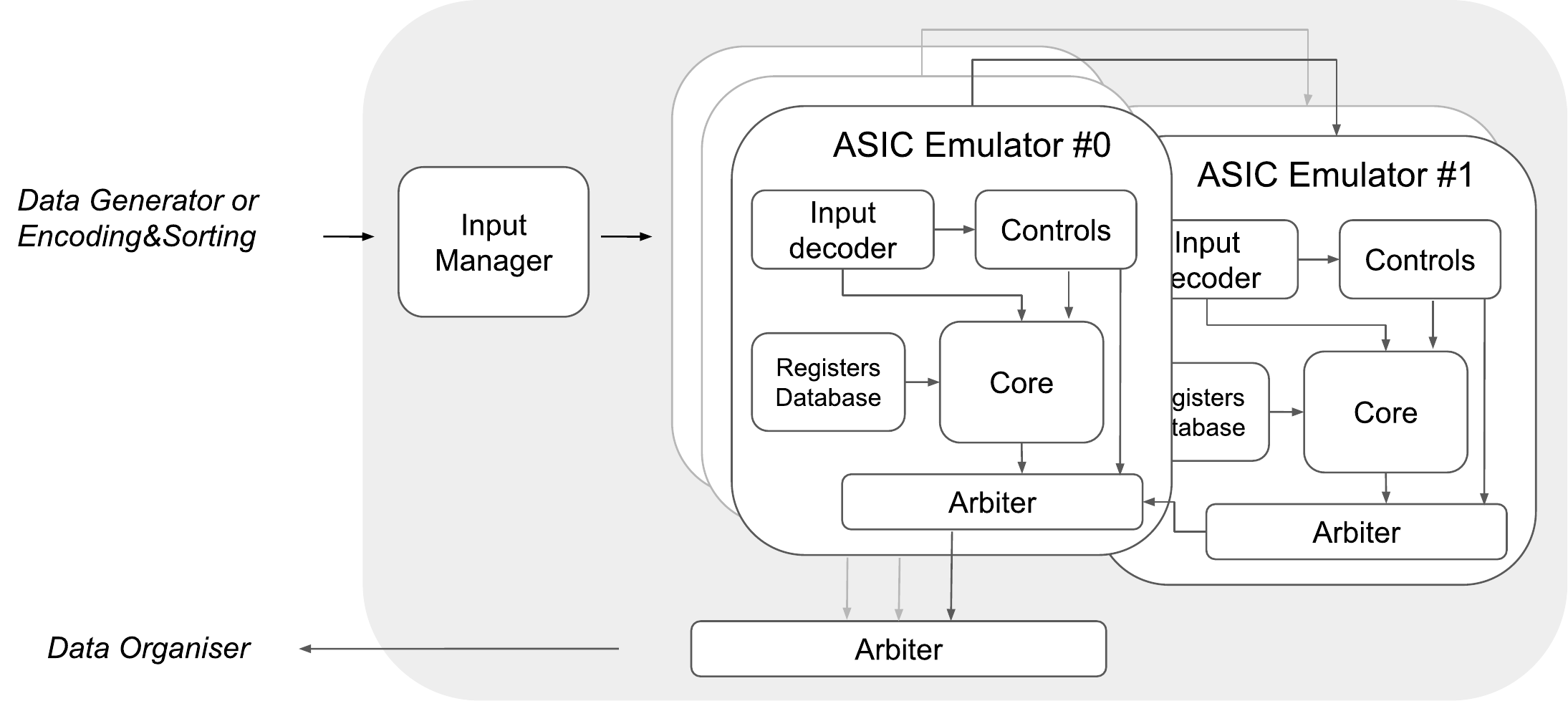}
 \caption{ASIC Emulator and Interface block diagram.}
 \label{fig:ASIC_Emu}
\end{figure}

The ASIC Emulator implements the AM pattern matching by storing 16 patterns per instance, each configurable at run-time.
It uses the same data and road bus input and output interfaces as the internal core of the AM ASIC~\cite{AM08}.
To fully mimic a single group of AM ASICs on the PRM, a daisy chain of five emulator instances is implemented.
Figure \ref{fig:ASIC_Emu} shows a block diagram of a group of five ASIC Emulators.

The input bus is designed to supply the ASIC Emulator with three types of data: cluster data, control commands, and the idle character.
The Input Decoder distributes cluster data to the Core block and commands to the Controls block.
The Core block stores the patterns and implements the pattern matching functionality. 
The Controls block decodes the commands and implements the execution timing structure.
Events are initialized before the first SSIDs are to be sent to the core for comparison, and ended to gather the information from all matches.
All matched roadIDs are then read out in series.
The roadIDs from the Core block are combined in the Arbiter block with the output from the Controls block.
The output data streams of the five emulators are connected in three daisy-chains of up to two emulators, mimicking the hardware implementation on the demonstrator.
The total output data is collected in the ASIC Interface via another Arbiter block and sent to the Data Organiser.

\subsection{Data Organiser}
\begin{figure}[ht]
 \centering
 \includegraphics[width=0.8\textwidth]{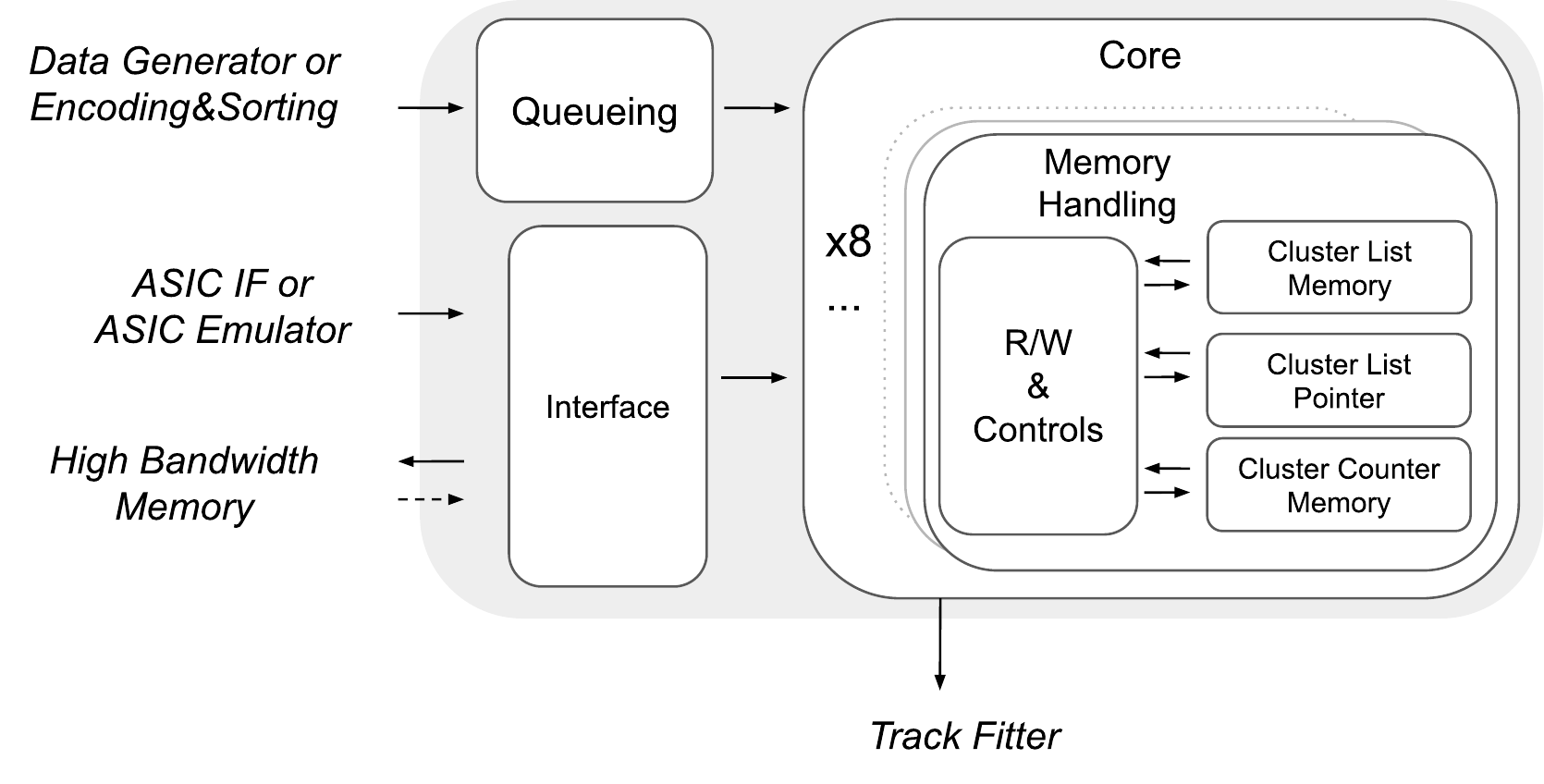}
 \caption{Data Organiser block diagram.}
 \label{fig:DOdiagram}
\end{figure}

The Data Organiser is an on-the-fly database that allows fast access to the cluster position and their SSID. This is critical to extract the cluster from a matched pattern for the linearized track fit.
A schematic view of the Data Organiser block is shown in Figure \ref{fig:DOdiagram}.

The clusters and SSIDs coming from the Data Generator or from the SSID Encoding and Sorting blocks, already organised per layer, are temporarily stored in FIFOs in the Queueing block.
This makes sure that the readout of the previous event has been completed.

Later, a three-memory structure is used in order to better retrieve the clusters.
The Data Organiser has been designed under the assumption that all clusters for a given SSID arrive in one uninterrupted sequence.
The Cluster List Memory sequentially stores each cluster received. 
The Cluster List Pointer  stores the Cluster List Memory address of the first cluster stored for each SSID. 
The Cluster Counter Memory stores the number of clusters in each SSID.
These functions are performed in parallel for all eight layers.

The Interface block forwards the roadID from the ASIC Emulator blocks towards the High Bandwidth Memory.
The SSID list returned by the HBM for a matched pattern is used to retrieve the correct clusters from the DO memories.
The found roads are then sent towards the Track Fitter block.

\subsection{High-Bandwidth Memory Interfaces}
\label{sec:HBM}
\begin{figure}[ht]
 \centering
 \includegraphics[width=0.8\textwidth]{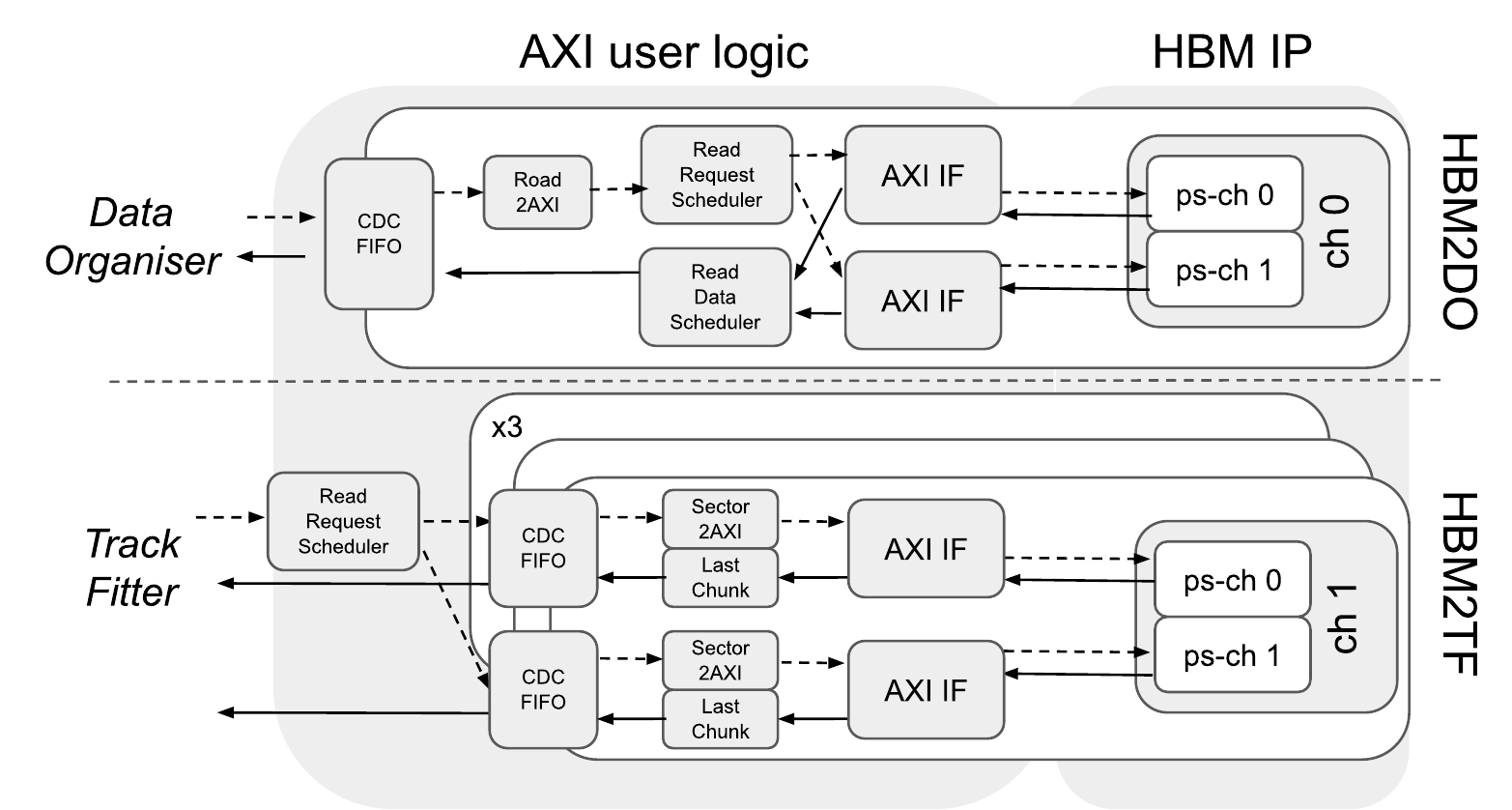}
 \caption{High-Bandwidth Memory user logic block diagram.
 CDC FIFO~=~Clock Domain Crossing FIFO, ps-ch~=~pseudo-channel.
 The functionality of the individual blocks is described in the text.}
 \label{fig:HBM}
\end{figure}

The High-Bandwidth Memory block, shown in Figure \ref{fig:HBM}, stores the patterns as well as the constants needed for the computation of the quality of the track fit ($\chi^2$) and the corresponding track parameters.

FIFO buffers are used at the interfaces between the core logic and the HBM AXI user logic for crossing the clock domains, as each HBM channel has an independent clock, and also for handling back pressure. 
Two distinct user interfaces are implemented to best match the required data rates and interfaces of the Data Organiser (HBM2DO) and the Track Fitter (HBM2TF).

As the individual data packets requested from the HBM are rather small (ranging from 18~Bytes for each pattern for the Data Organiser, to a maximum of 364 Bytes for each set of constants for the Track Fitter), all channels used of the Intel HBM IP~\cite{HBM} are configured in burst length 4 mode, returning 32 Bytes per single read request.
The HBM IP internal Read-Order-Buffers are enabled to simplify the data flow.
They can be bypassed to improve latency and bandwidth. 
The remaining settings of the IP have been optimized in simulation studies for lowest latency and highest throughput.
Latency histograms have been implemented within the AXI interfaces to be able to monitor the read data turnaround time.

The Data Organiser requests the patterns from the HBM2DO interface by sending the roadID paired with a requestID. 
The HBM2DO translates the roadID into a valid memory address (Road2AXI).
Two pseudo-channels are used within the HBM2DO to match the maximum rate requirements.
A scheduler checks which pseudo-channel can serve a read request and evens the load on the two pseudo-channels connected using round-robin arbitration.
The requestID is stored temporarily matching the AXI-ID used to send out the read request to the HBM IP.

The data received from the HBM is matched to the requestID retrieved via the AXI-ID.
A scheduler merges the data streams of the two pseudo-channels towards the output FIFO.
The data is split into sectorID and SSIDs for the Data Organiser to continue processing.

The Track Fitter requests constants from the HBM2TF for $\chi^2$ and/or parameter computations specifying the sectorID.
The request gets distributed to one of 6 pseudo-channels connected using round-robin arbitration, taking into account if a pseudo-channel is currently busy.
Within a pseudo-channel interface, a number of consecutive AXI memory read requests is issued to retrieve a full set of constants (Sector2AXI).
Upon return of the data, the last chunk of a set of constants is flagged which allows the Track Fitter to easily identify when a set of constants has been fully retrieved.
This allows operation of the HBM even without the Read-Order-Buffers being enabled.

In order to write the patterns and constants data into the HBM, an IPbus~\cite{IPbusref} compatible interface has been implemented. 
For online monitoring and configuring of the HBM, an IPbus to Advanced Peripheral Bus (APB) translator interface has been developed.

\subsection{Track Fitter}
\begin{figure}[ht]
    \centering
    \includegraphics[width=0.8\textwidth]{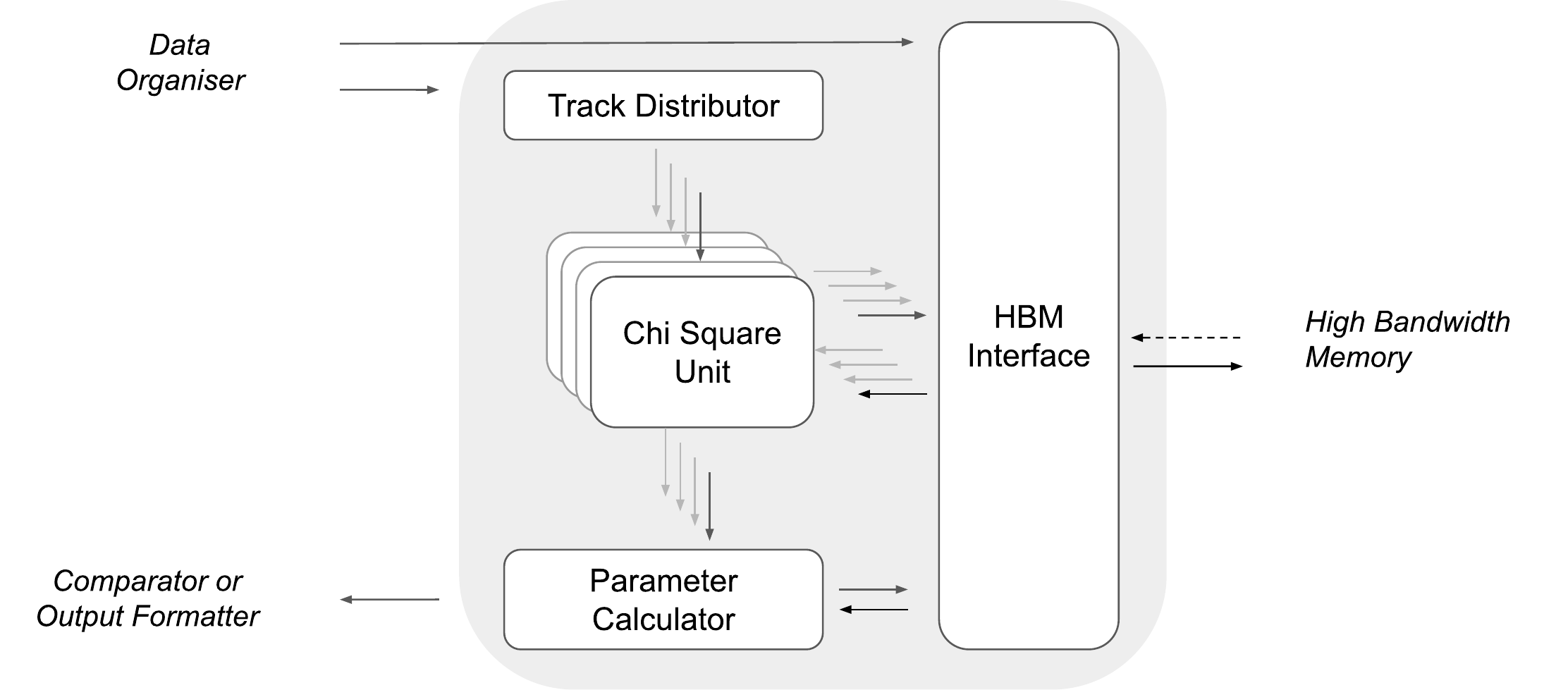}
    \caption{Flow diagram of the Track-Fitter. Buffers and other minor entities are not displayed.
    \label{Fig_TF_flow}}
\end{figure}

The Track Fitter block, shown in Figure \ref{Fig_TF_flow}, assesses the quality of the track candidates by computing the $\chi^2$ value, and then extracts the track parameters for those tracks that pass the $\chi^2$ test.
Upon arrival of a road data set, the Track Fitter retrieves the constants for the $\chi^2$ computation from the HBM.
All constants are encoded as signed \SI{32}{bit} floating point numbers. 
Hence, all DSPs implemented are used in floating point mode.

In the Track Distributor, all possible combinations of clusters from the matched patterns are generated as candidate tracks for the fit.
To match the required data rates, this is performed through four parallel sub-entities, each associated with a Chi Square Unit block.
For each track candidate, the HBM interface retrieves the corresponding constants. 
All constants requests are buffered while one set of constants is retrieved, and a supporting entity aligns each set to the corresponding track data.

The Chi Square Unit receives a track candidate ($x_j$), and its IDs and constants ($S_{i,j}$ and $h_i$) simultaneously.
Here, the scalar products per number of coordinates are executed in parallel for all degrees of freedom, according to the inner sum of Equation~\ref{eq:chi}, using a cascaded pipeline. 
Likewise, the scalar product over all degrees of freedom is performed, according to the outer sum of the Equation~\ref{eq:chi}. 
This design can perform a calculation at every clock cycle with a fixed latency, and originates from the works of \cite{braagaard} and \cite{jensen}.
Finally, the track parameters for this track candidate will be computed if the resulting $\chi^2$ is below a predefined threshold. 

Since only \SI{8}{\%} of the track candidates are expected to pass the selection in the studied configuration, see Table~\ref{tab:PRM_parameters}, only one Parameter Calculator block is instantiated. 
Here, the HBM interface provides again the constants, and a scalar product per parameter is executed according to Equation~\ref{eq:param}.
The resulting parameters, the $\chi^2$ of the track fit, and the clusters are converted into \SI{16}{bit} fixed point numbers and transferred to the comparator block of the data generator.

\subsection{Monitoring}
\label{subsec:mon}

\begin{figure}[ht]
 \centering
 \includegraphics[width=0.8\textwidth]{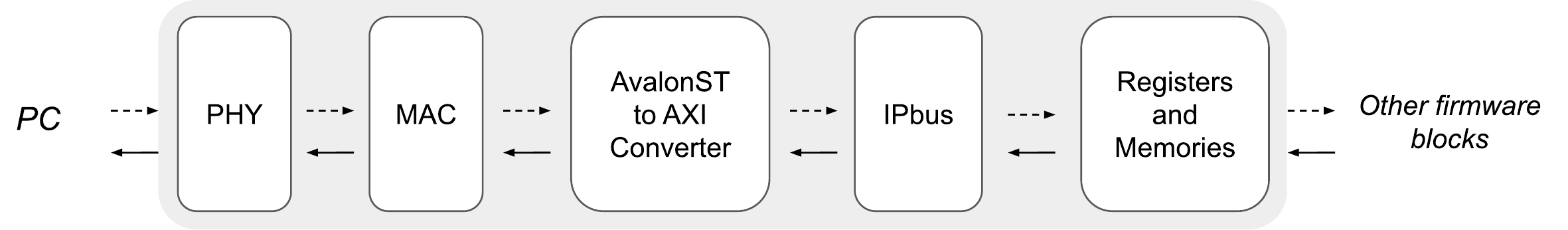}
 \caption{Monitoring block diagram.}
 \label{fig:MonDiagram}
\end{figure}

The Monitoring block handles the communication between the PRM FPGA and an external PC, via Ethernet using IPbus protocol over UDP.
The block diagram is shown in Figure \ref{fig:MonDiagram}.

The Physical (PHY) and the Medium Access Control (MAC) layer blocks are Ethernet-dedicated IP cores provided by Intel. 
To allow the communication between the MAC block (Avalon ST interface) and the IPbus block (AXI Interface), a dedicated converter module has been designed. 
Information about the functionality of the IPbus core can be found in~\cite{IPbusref}.  
Hardware checks (e.g. voltages, temperatures), firmware configuration (e.g. data generator setup), and data flow monitoring (e.g. FIFO full and empty, counters) can be all executed via this block.
The software part of the communication is provided by the same group which developed the IPbus core \cite{IPbusref}.


\section{Implementation and results}
\label{sec:details}

The firmware described above has been implemented and tested on an Intel Stratix 10 MX development kit, which features a nearly-identical FPGA to the one chosen for the PRM prototype\footnote{The FPGA on the development kit (part number 1SM21BHU2F53E1VG) and the one on the PRM (part number 1SM21BHN3F53E3VG) differ only in transceiver count, transceiver speed grade, and FPGA fabric speed grade.}.
Simulation studies have been conducted using QuestaSim version 2019.4 \cite{questaref}.
The Intel Quartus Prime Pro software version 19.2 \cite{quartusref} has been used for hardware synthesis and fitting.
The results of the initial firmware implementation are presented. 
It does not include further optimizations that would allow to meet all the requirements of the HTT system; 
these are presented as proposals in the final section of the paper.

\subsection{Resource usage}
The resource utilization after the final fitting step for each entity of the standalone implementation is presented in Table~\ref{tab:FWres}. 
The block-by-block as well as the total occupancy are presented in absolute numbers and in \% fraction of the full device resources. 
In addition to the individual components, the resource usage of the top level (where all blocks are interconnected) is shown with the addition of SignalTap~\cite{stpref} debug monitoring.
All blocks that are expected to consume significant logic resources are implemented and considered in this measurement.
More advanced features, in particular the extension to include tracks with missing clusters, are not implemented.


\begin{table}[h]
\begin{center}
  \begin{adjustbox}{width=.9\textwidth}
    \begin{tabular}{@{}lrrrrr@{}}
\toprule
\textbf{}&
\multicolumn{1}{c}{\textbf{\begin{tabular}[c]{@{}r@{}}ALMs\\k\hspace{.5em}(\%)\end{tabular}}}&
\multicolumn{1}{c}{\textbf{\begin{tabular}[c]{@{}r@{}}Logic Registers\\k\hspace{1em} (\%)\end{tabular}}}&
\multicolumn{1}{c}{\textbf{\begin{tabular}[c]{@{}r@{}}Block Memory\\kb\hspace{1em}(\%)\end{tabular}}}&
\multicolumn{1}{c}{\textbf{\begin{tabular}[c]{@{}r@{}} \hspace{.5em}DSPs\hspace{.5em} \\\#\hspace{.1em} (\%)\end{tabular}}}&
\multicolumn{1}{c}{\textbf{\begin{tabular}[c]{@{}r@{}} \hspace{.5em}HBM\hspace{.5em} \\GB\hspace{.1em} (\%)\end{tabular}}}\\
\midrule
Data Generator        & 17.36 (2.47)   & 26.67 (0.95)  & 206.03 (0.15)    & 0 (0.00)   & 0 (0.00) \\
ASIC Emulator         & 7.06 (1.08)    & 16.42 (0.58)  & 45.28 (0.03)     & 0 (0.00)   & 0 (0.00) \\
Data Organiser        & 4.78 (0.68)    & 6.20 (0.22)   & 11,625.48 (8.29) & 0 (0.00)   & 0 (0.00) \\
High Bandwidth Memory & 14.15 (2.01)   & 39.85 (1.42)  & 188.91 (0.13)  & 0 (0.00)   & 2.00 (25.00) \\
Track Fitter          & 112.06 (15.95) & 235.68 (8.38) & 541.84 (0.39)    & 424 (10.7) & 0 (0.00) \\
Monitoring            & 7.90 (1.12)    & 12.97 (0.46)  & 934.15 (0.67)    & 0 (0.00)   & 0 (0.00) \\
Top level + Debug     & 8.19 (1.17)    & 17.14 (0.61)  & 321.06 (0.23)    & 0 (0.00)   & 0 (0.00) \\
\midrule
\textbf{TOTAL} &\textbf{172.04 (24.48)} &  \textbf{354.92 (12.63)} &\textbf{13,862.75 (9.89)} & \textbf{424 (10.71)} &  \textbf{2.00 (25.0)} \\ 
\midrule
\textbf{\begin{tabular}[c]{@{}l@{}}TOTAL w/o Data Generator \\  and ASIC Emulator \end{tabular}}   &  \textbf{154.69 (20.93)} &  \textbf{328.25 (11.09)} &\textbf{13,611.44 (9.71)} & \textbf{424 (10.71)} &  \textbf{2.00 (25.0)} \\ 
\bottomrule

    \end{tabular}
  \end{adjustbox}
  \caption{Resource usage of the PRM standalone firmware on the Intel Stratix 10 MX 2100. 
  Absolute values are given (in multiples of 1000 = k), accompanied by the corresponding percentage of the device in brackets. For considerations on the final design the Data Generator and the ASIC Emulator blocks should not be taken into account, as they are used only for standalone testing.
  Abbreviations are explained in the text.
  }
\label{tab:FWres}
\end{center}
\end{table}


The Track Fitter, where the track-dedicated algorithms are implemented, is the block using the most logic resources, i.\,e.\, Adaptive Logic Modules (ALMs), registers, and Digital Signal Processing (DSP) blocks. 
The Data Organiser, where the data are stored during the pattern recognition step, is the block consuming the majority of the memory resources.
The FPGA occupancy does not exceed one quarter of the available resources.
Considering that optimizations could further reduce the resource utilization significantly, it would be feasible to fulfill the PRM system requirement of implementing four independent instances in parallel, including the missing extensions.
For operation on the PRM board, the ASIC emulator would be replaced by the actual ASIC interface and the Data Generator by the input/output buffers, synchronization modules\footnote{The Cluster to SSID Encoding block is already included in the Data Generator.}, and the output formatter.
Hence, the total resource utilization excluding the contribution of these two entities should be considered.


\subsection{Clock and power studies}
\label{sec:clk}
In order to satisfy the system requirements presented in Table \ref{tab:PRM_parameters}, the PRM firmware was planned with the goal of running the Track Fitter block at \SI{200}{MHz}, and the rest of the logic at \SI{250}{MHz}.

The hardware tests with the standalone firmware have been performed constraining the design for a core clock speed of \SI{100}{MHz}, aiming to verify the hardware implementation rather than optimizing for highest processing speed.
A single clock domain has been used for the entire design, with the exception of the HBM that is operated at a memory clock frequency of \SI{600}{MHz} and a core clock frequency of \SI{300}{MHz}.
Without violating internal setup and hold time requirements, the maximum core clock frequency $f_{\rm{max}}$ reported by the Timing Analyzer (part of the Intel Quartus Prime Pro software) is $f_{\rm{max}} = \SI{105}{MHz}$ in the worst case (Slow \SI{900}{mV} Model @ \SI{100}{^\circ C}).

Under these conditions, the Quartus Power Analyzer tool estimated the total thermal power dissipation to be about 11 W, where about 6 W are due to the device static thermal power dissipation. 
Unfortunately, the Quartus tool does not include a power model for the HBM component.
Its power consumption can hence not be properly estimated in this stage of the development. 

In order to better understand the most critical blocks, dedicated \SI{100}{MHz} clock lines are used for each block by splitting the core clock with a PLL. 
The results per block are reported below in Table~\ref{fig:Timing_models}.
It is important to note that in these studies 
clock domain crossings have not been properly implemented.
Although this should not affect the performance and functionality of the individual blocks, treating these crossings correctly is important for their interconnection, and could potentially further improve the results.
The following operating conditions are used for this study.
The $V_{cc}$ minimum supply voltage is fixed at \SI{900}{mV}, and the temperature is set to \SI{0}{^{\circ}C} and \SI{100}{^{\circ}C}, which represent the two extreme corner cases for the $f_{\rm{max}}$ value.

\begin{table}[h]
  \centering
\resizebox{\textwidth}{!}{\begin{tabular}{ccccc>{\color{gray}}c>{\color{gray}}c}
\textbf{\begin{tabular}[c]{@{}c@{}}Slow 900 mV\\ Model\end{tabular}}  & \textbf{\begin{tabular}[c]{@{}c@{}}Target Clock\\ Speed (MHz)\end{tabular}} &  \textbf{\begin{tabular}[c]{@{}c@{}}Data Organiser \\ $f_{\rm{max}}$ (MHz)\end{tabular}} & \textbf{\begin{tabular}[c]{@{}c@{}}HBM\\ $f_{\rm{max}}$ (MHz)\end{tabular}} & \textbf{\begin{tabular}[c]{@{}c@{}}Track Fitter\\ $f_{\rm{max}}$ (MHz)\end{tabular}} &
\textbf{\begin{tabular}[c]{@{}c@{}}Data Generator\\ $f_{\rm{max}}$ (MHz)\end{tabular}} & \textbf{\begin{tabular}[c]{@{}c@{}}ASIC Emulator\\ $f_{\rm{max}}$ (MHz)\end{tabular}} \\

\hline
\textbf{100 $^{\circ}$C } & \multirow{2}{*}{100}   & 141.60 & 301.39 & 114.94 & 141.26 & 136.69\\
\textbf{0 $^{\circ}$C} &                           & 154.63 & 322.37 & 125.20 & 152.39 & 148.08 \\


\hline

\end{tabular}}
\caption{Per block $f_{\rm{max}}$ derived from static timing analysis for a target clock of \SI{100}{MHz} (\SI{300}{MHz} for the HBM). For considerations of the final design only the Data Organiser, HBM and Track Fitter should be taken into account.}
\label{fig:Timing_models}
\end{table}



A test has been made to evaluate the work needed to fulfill the HTT system requirement to run at higher frequencies.
With the current design constraints and targeting an input clock speed of \SI{200}{MHz}, an $f_{\rm{max}} = \SI{174}{MHz}$ (Slow \SI{900}{mV} Model @ \SI{100}{^\circ C}) was achieved, not yet meeting our requirements. 
The Track Fitter block yields the lowest $f_{\rm{max}}$ of all the individual blocks. 
An algorithm rework would be required for a substantial speed optimization, as will be discussed in the final section.

For this clock speed, the Quartus Power Analyzer tool estimated the total thermal power dissipation to be about \SI{15}{W}, where about \SI{6}{W} are due to the device static thermal power dissipation, excluding the HBM contribution.







\subsection{Performance studies}
\label{sec:perf}
The performance of the standalone firmware, with a focus on the processing latency, has been evaluated in simulation using dedicated test vectors derived from the software simulation of the HTT system~\cite{TDAQ_amend}.
A test vector is a collection of information of cluster positions and track parameters, generated by the simulation that reproduces the algorithms of the linearized track fit, and converted into \SI{32}{bit} words readable by the Data Generator interface.
Simulated events with single muon tracks crossing the ITk layers are used.

Larger event sizes have been studied by repeating the same track multiple times as an intermediate step towards more realistic conditions. Each track consists of one cluster per layer. This was done to study the constants request rate and latency, as every track implies a request of constants towards the HBM.
Test vectors with few additional fake clusters, i.\,e.\, clusters not belonging to any real track, have also been used during hardware verification, but have not been part of the performance evaluation.
Studies trying to mimic realistic HL-LHC conditions by injecting events which fulfill approximately the number of roads and fits per event, as specified in Table~\ref{tab:PRM_parameters}, have been started but could not be concluded for this publication.
For the performance studies reported here, the SSID Encoding and Sorting block within the Data Generator has not been included.
However, both Data Generator modes, with and without the SSID Encoding and Sorting block, have been verified in hardware.
Despite the limitations in clock speed seen in the hardware implementation, the simulation has been run at a clock speed of \SI{250}{MHz}, with the exception of the HBM user logic running at \SI{300}{MHz}, as explained in section~\ref{sec:clk}.
\begin{figure}[h]
    \centering
    \begin{subfigure}[b]{0.45\textwidth}
        \includegraphics[width=\textwidth]{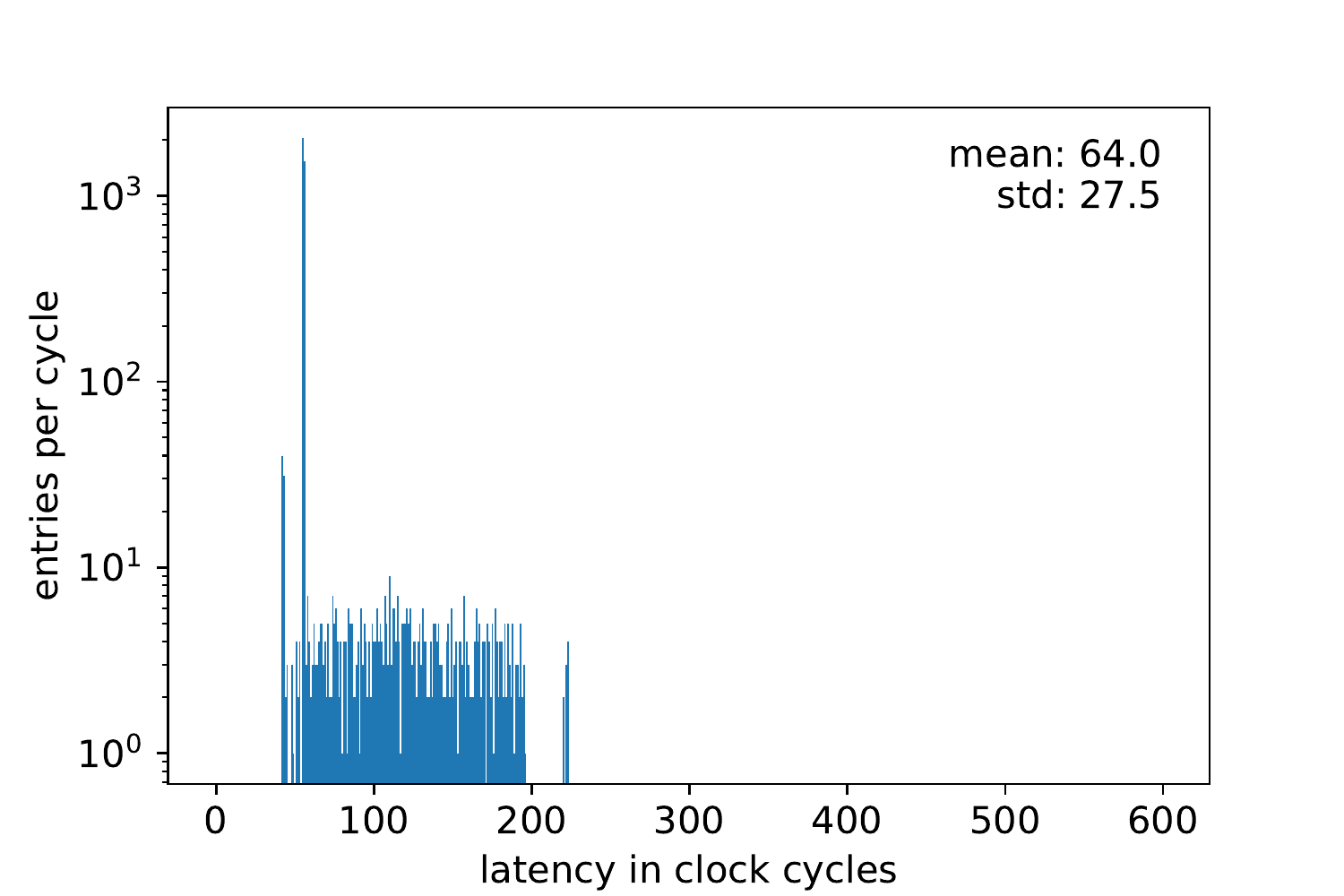}
        \caption{HBM pattern read latency, 1 track/event}
        \label{fig:hbm2do_1track}
    \end{subfigure}
    \begin{subfigure}[b]{0.45\textwidth}
        \includegraphics[width=\textwidth]{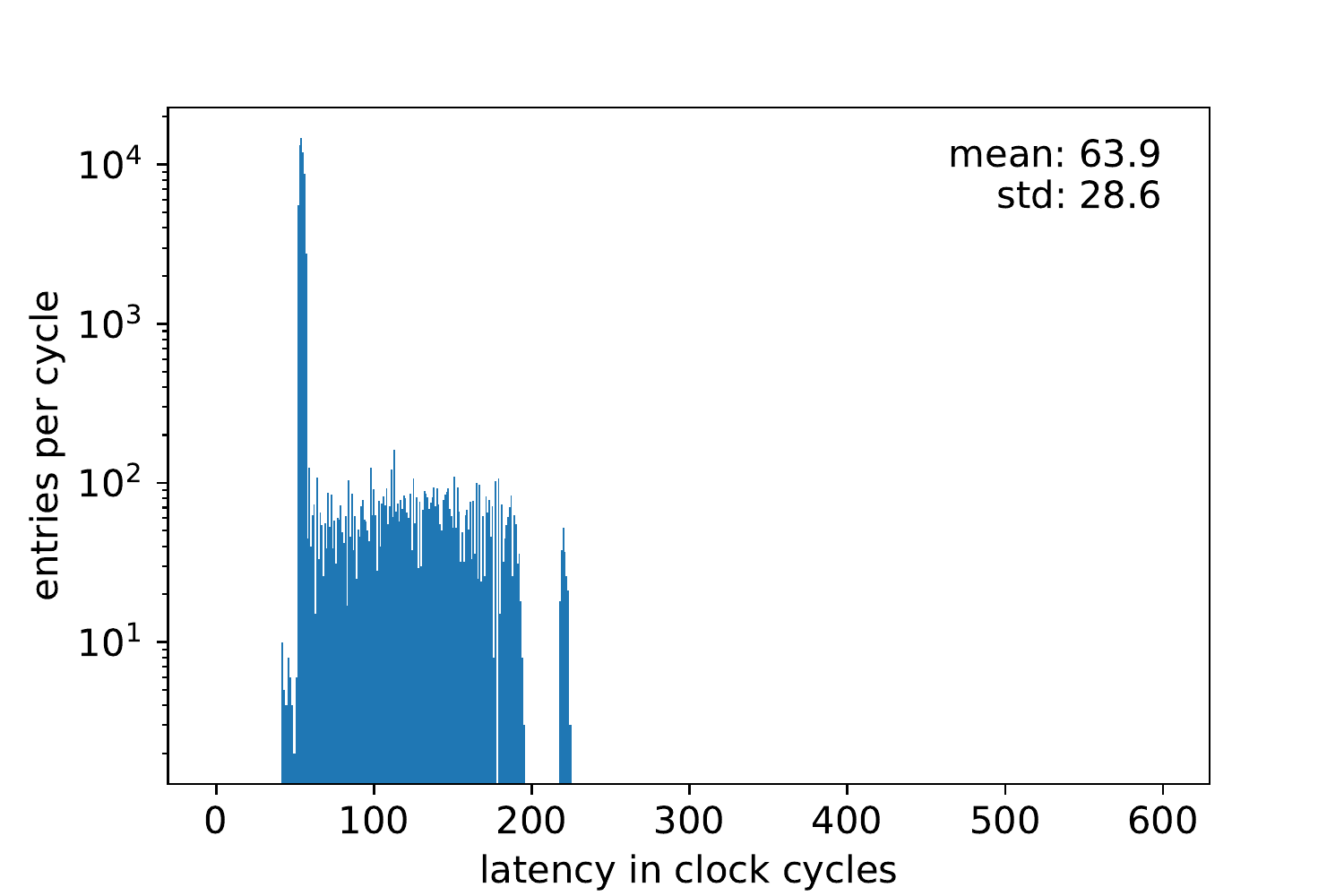}
        \caption{HBM pattern read latency, 16 tracks/event}
        \label{fig:hbm2do_16tracks}
    \end{subfigure}
    \begin{subfigure}[b]{0.45\textwidth}
        \includegraphics[width=\textwidth]{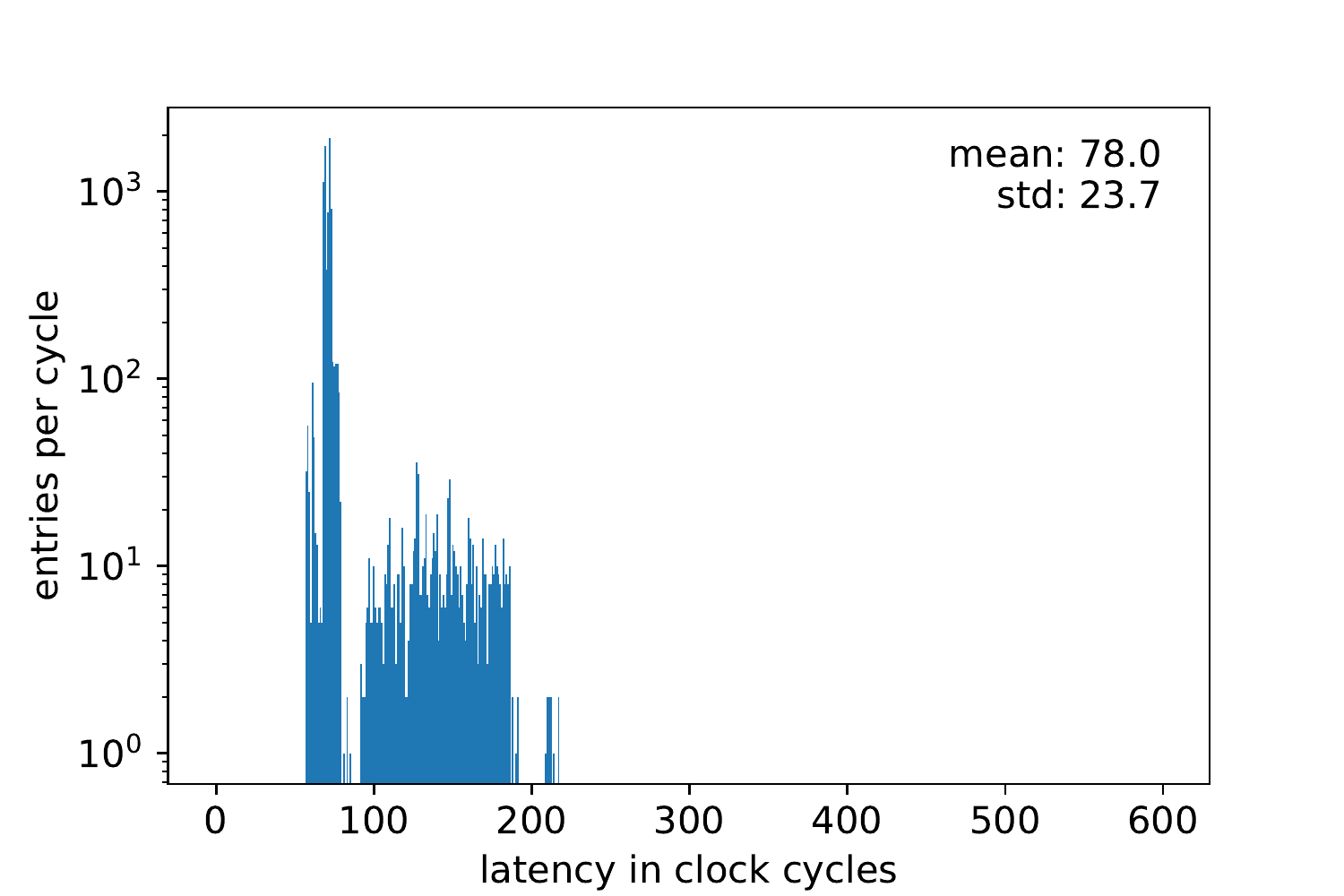}
        \caption{HBM constants read latency, 1 track/event}
        \label{fig:hbm2tf_1track}
    \end{subfigure}
        \begin{subfigure}[b]{0.45\textwidth}
        \includegraphics[width=\textwidth]{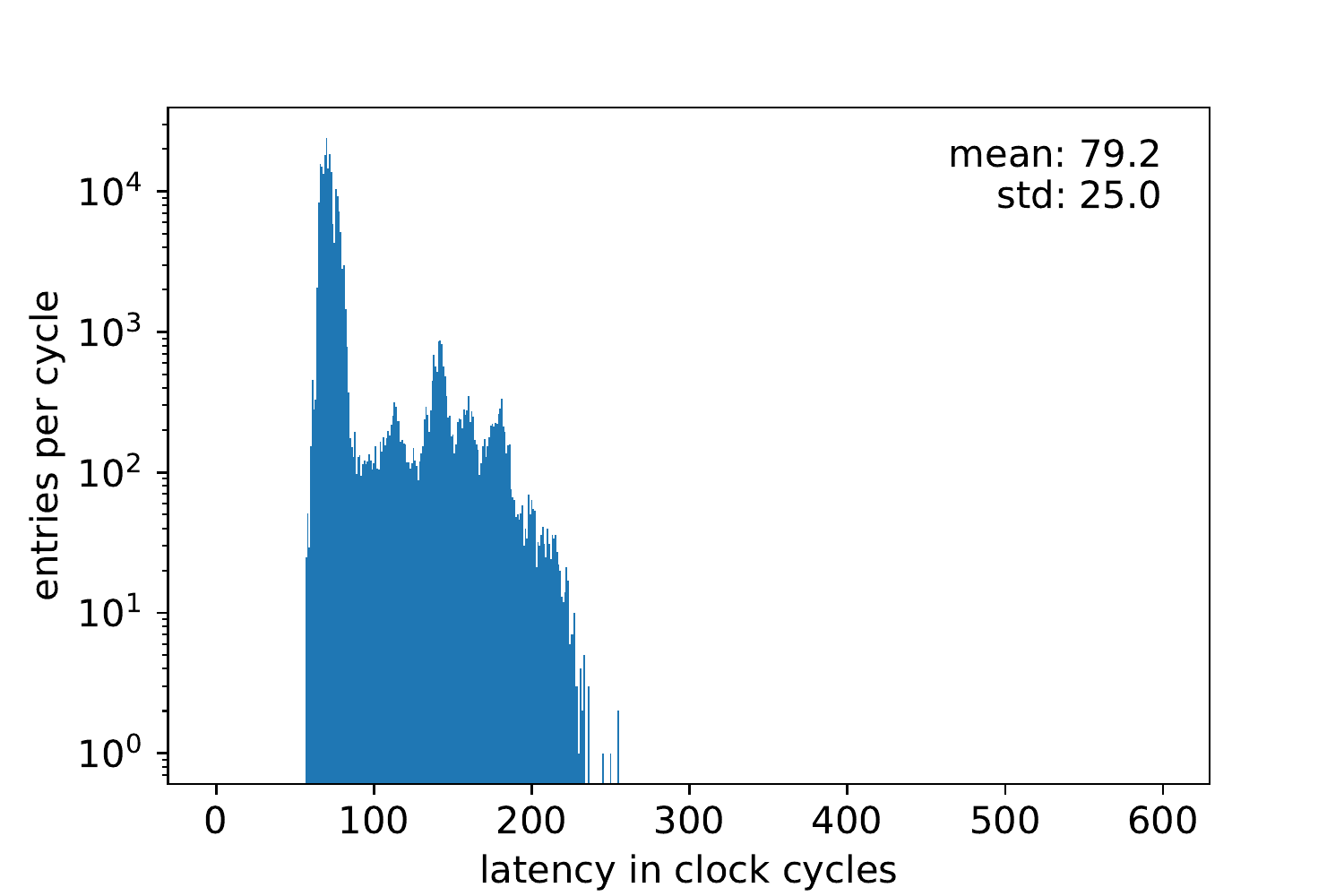}
        \caption{HBM constants read latency, 16 tracks/event}
        \label{fig:hbm2tf_16tracks}
    \end{subfigure}
    \begin{subfigure}[b]{0.45\textwidth}
        \includegraphics[width=\textwidth]{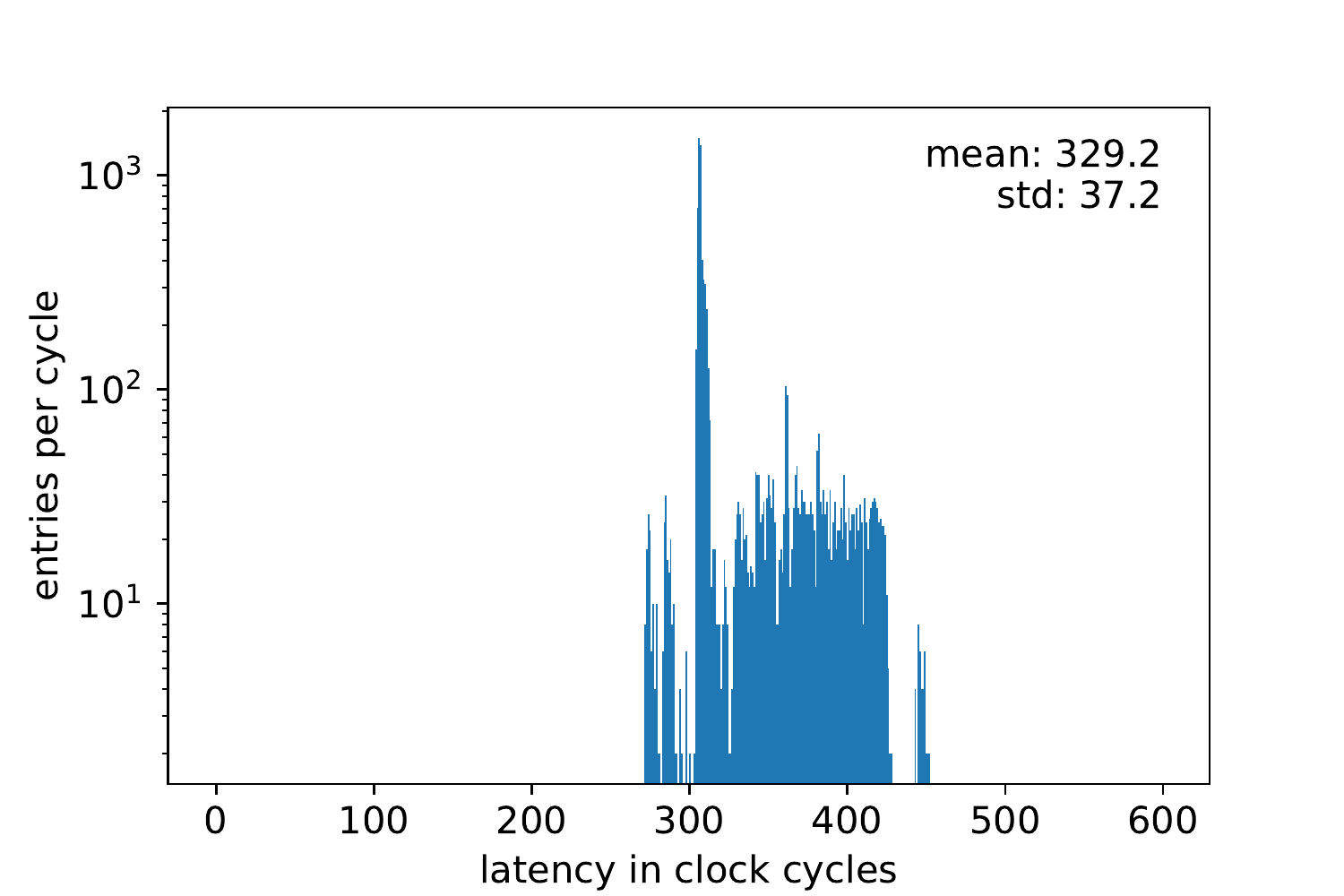}  
        \caption{Track latency, 1 track/event}
        \label{fig:track_latency_1track}
    \end{subfigure}
     \begin{subfigure}[b]{0.45\textwidth}
        \includegraphics[width=\textwidth]{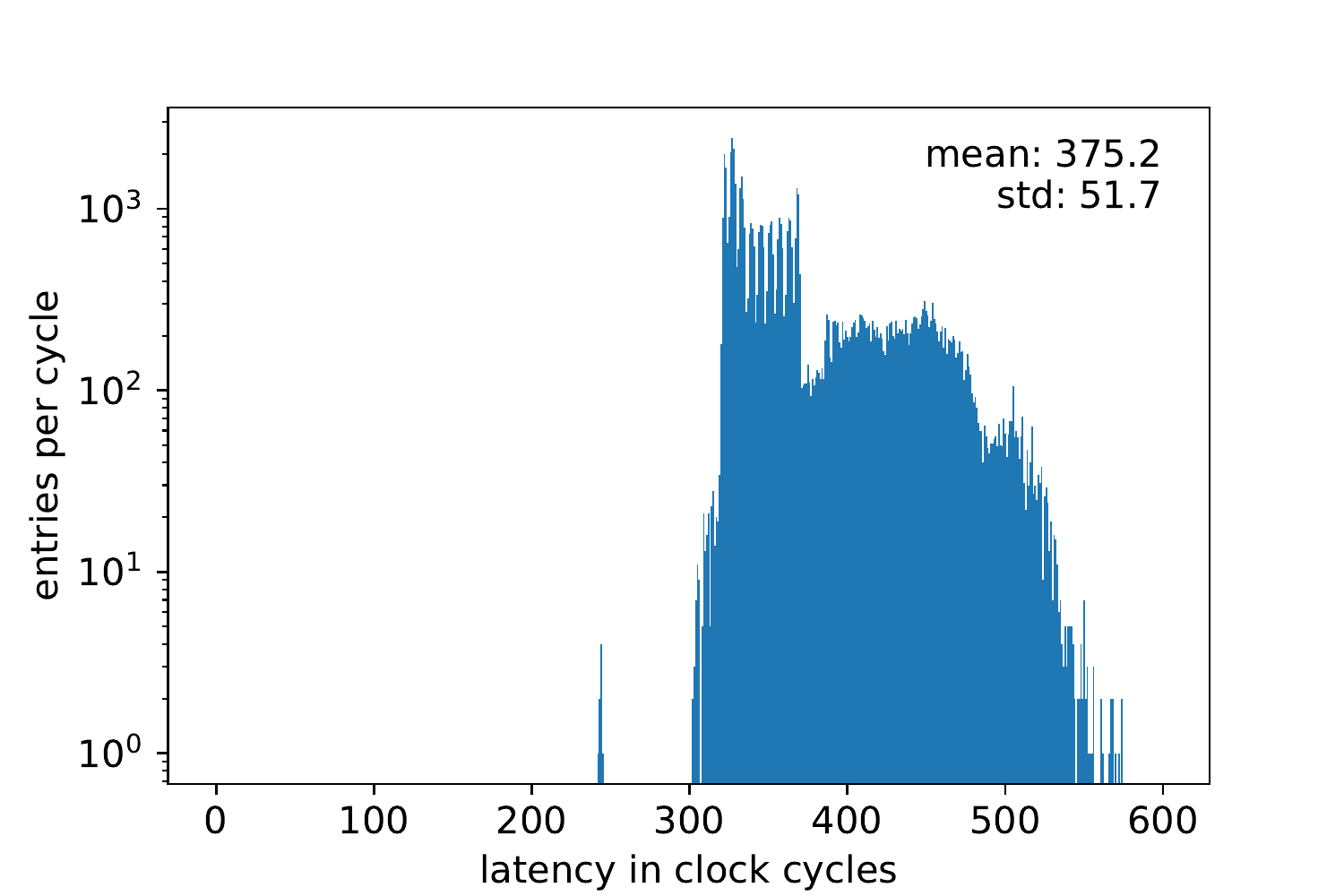}  
        \caption{Track latency, 16 tracks/event}
        \label{fig:track_latency_16tracks}
    \end{subfigure}
    \caption{Processing latencies of different parts of the PRM firmware for events with a single track (left) and events with 16 tracks (right) in clock cycles of $f_{core} = \SI{250}{MHz}$.
    (a) and (b) show the latency of patterns being returned from the HBM when requested by the Data Organiser.
    (c) and (d) show the latency of full sets of constants returned from the HBM when requested by the Track Fitter.
    (e) and (f) show the latency of a track through the processing pipeline, between its appearance as a road at the output of the ASIC emulator and the final track at the output of the Track Fitter.
    For all distributions, mean and standard deviations are given.}
    \label{fig:latencies}
\end{figure}

For events with single and multiple tracks, the processing latency per block has been measured as a function of the event rate.
Single-track events are filled at a rate of \SI{886}{kHz}.
For events with 16 tracks we chose an event rate of \SI{541}{kHz}, leading to road, track, cluster, constants readout and fit rates of \SI{8.65}{MHz}.
These event rates are chosen to keep well separated single-event processing, since the handling of multiple concurrent events is not implemented.

Figure~\ref{fig:latencies} shows the histograms of the memory latency between the HBM and the Data Organiser, between the HBM and the Track Fitter, and also the full processing latency of each individual track from the output of the ASIC emulator to the output of the Track Fitter.
About 4000 events have been processed to capture the effect of the HBM on the latency.
The majority of the read requests are served in a short time of about \SIrange{250}{300}{ns}; 
there are tails of up to about \SI{1}{\micro s} due to the dynamic nature of the HBM, being a DRAM, which requires refreshing of the stored data every few microseconds.
During the refresh, the HBM pseudo-channel cannot serve any read requests.
This behaviour is inherently different compared to internal/external SRAM or the FPGA's internal block RAM.
Compared to single track events, no significant increase on the average HBM latency per pattern or set of constants is observed for events with 16 tracks, because the overall HBM request rates are still well below the bandwidth limits. 

The accumulated HBM latency accounts for the biggest part of the total latency as there are three separate accesses per track: to retrieve the pattern, the constants for the $\chi^2$ computation, and the constants for the track parameter calculation.
Considering the mean values of the latency distributions, about $\SI{66}{\%}$ ($\SI{60}{\%}$) of the latency is due to the HBM for events with single tracks (16 tracks).
Moreover, we have an increase of latency in the total processing pipeline.
This can be linked to the Data Organiser, which in its current state needs 3 clock cycles to transfer a single road to the Track Fitter, and to the Track Fitter itself, whose processing can be temporarily interrupted when the requested constants are delayed.
The average total processing latency is \SI{1.3}{\micro s} (\SI{1.5}{\micro s}) for events with single tracks (16 tracks), with increasing tails in the distribution for higher track rates\footnote{In the presence of high energy jets and pileup 200 we would expect an average of about 320 roads and 80 tracks per event per group of ASICS.}. 
These tails could be reduced by optimizing the design of the Track Fitter and the output of the Data Organizer. 
No significant increase in latency due to the HBM is expected at the required rates (Table~\ref{tab:PRM_parameters}).
More details on design optimizations are given in the following section.

\subsection{Considerations for a final design}
\label{sec:consid}



These studies show a possible implementation of the main functionalities of the firmware for a linearized track fit with 8 detector layers, associated to AM pattern matching.
Not all requirements from the HTT system design, outlined in Table~\ref{tab:PRM_parameters}, are achieved at this stage of our development.
For example, the $\chi^2$ fit rate requirement of \SI{714}{MHz} can be achieved with a clock speed for the Track Fitter of \SI{200}{MHz}, but hit a limit of $f_{\rm{max}} = \SI{175}{MHz}$ in our design. 
Because of this, and in order to lower its resource usage, the Track Fitter block would be the very first candidate for optimizations.
This is despite the expectation that it will consume the largest fraction of resources compared to all other blocks.
For instance, the interface between the Track Fitter and the HBM could be simplified such that there is one dedicated Chi Square Unit per HBM pseudo-channel, instead of the current implementation where all four Chi Square Units can retrieve their constants from all of the six HBM pseudo-channels.
This may have a positive impact on the achievable $f_{\rm{max}}$ within the Track Fitter, and therefore increase the overall throughput.
The mean processing latency measured in simulations exceeds the target of \SI{1}{\micro s} for the low-latency scenario.
Several options could be adopted to reduce latency:
\begin{itemize}
    \item From simulation studies we know that the Read-Order-Buffer within the HBM Controller yields a large latency penalty, especially when considering higher request rates.
Removing this buffer, however, comes at the cost of increasing complexity of the data processing at the Track Fitter and Data Organiser.
    \item The constants for the $\chi^2$ and parameter calculations within the Track Fitter could be requested in parallel, as opposed to the current implementation where the second set of constants is only requested when the track has passed the $\chi^2$ cut.
This comes at the cost of increased HBM bandwidth.
    \item The constants for the Track Fitter could be requested as soon as the HBM has retrieved the pattern for the Data Organiser in parallel to the processing pipeline.
    This is instead of waiting for the data being transferred from the Data Organiser to the Track Fitter, which then requests the constants. Moreover, the sectorIDs could be stored in the FPGAs internal eSRAM, which has a much shorter read latency than the HBM. That would allow for truly parallel retrieval of the patterns and the fit constants from the HBM.
    From simulation studies of the HBM model at the required roads processing rates and constants readout rates specified in Table~\ref{tab:PRM_parameters}, it is expected that \SI{99}{\%} of patterns and constants would be returned within \SI{650}{ns} after their request.
    \item The Track Fitter could be adapted to start calculations as soon as a single chunk of constants has arrived, instead of waiting for the full set of constants being returned from the HBM.
    This would however require a drastic re-design of the Track Fitter block.
    \item The overall number of constants could be reduced by adopting an algorithm as it is reported in~\cite{CMS_fitter}.
    This makes use of hit position transformations to address the non-linearities of the system in order to reduce the number of sectors.
    This could allow the removal of the need for an HBM to retrieve the constants at all.
    However, this would not only require a complete redesign of the blocks following the Data Organiser within the PRM firmware, but also major changes to the simulation software of the full HTT system.
    \item Lastly, clock speed may be increased if a better speed grade version of the FPGA is used. 
    This change would also impact the HBM's read latency, as the core and memory frequency could be raised as well.
    This change would not require any design changes on the hardware or firmware, but increase the board's production costs.
\end{itemize}

\section{Conclusion}

Within the HTT project, a custom-hardware tracking system has been designed.
It is relying on associative memories for pattern matching and FPGAs for data organization, communication, and track fitting.
Even if the project is abandoned in ATLAS, many studies have been performed to assess its feasibility, both on hardware demonstrators and with simulations.
The Pattern Recognition Mezzanine as part of the HTT system features the pattern matching and track fitting capabilities.
Key challenges to the design of the PRM firmware are the high processing rates, e.\,g.\, the fit rate of $\num{4}\times\SI{714}{MHz}$, and the low-latency target of order \SI{1}{\micro s} requested for a short-latency L1 trigger scenario. 

This paper describes the implementation of the firmware for the PRM board, its validation in simulation, and tests on an Intel Stratix 10 MX development kit.
The major processing blocks, i.\,e.\,Data Organiser, Track Fitter, and HBM interfaces, have been implemented and tested using an AM ASIC Emulator and Data Generator developed specifically for this task.
Resource utilization of these blocks is in agreement with the requirements for a full PRM board.
Even if the requirements imposed by the HTT system design are not met, its feasibility with modern FPGAs with High-Bandwidth Memories is assessed.
Although High-Bandwidth Memories are optimized for highest throughput at a cost of higher latencies compared to SRAM or internal block memory, they can be utilized in low-latency ($\mathcal{O}(\SI{1}{\micro s})$) applications that require large data storage.
Many improvements are possible that could reduce the latency and resource utilization, leaving space for optimizations for custom hardware systems in future experiments.


\acknowledgments
We acknowledge support by the Federal Ministry of Research and Education (BMBF), the Swiss National Science Foundation (SNSF), the University of Geneva, Department of Nuclear and Particle Physics, the Italian Istituto Nazionale di Fisica Nucleare and the Aristotle University of Thessaloniki, School of Physics.
We would like to thank Francesca Pastore, Naoki Kimura, and Mel Shochet for valuable feedback that helped shape this article.


\end{document}